\documentclass[12pt]{article}
\usepackage{graphics}
\usepackage[dvips]{graphicx}
\usepackage{mathrsfs}
\usepackage{amssymb}
\usepackage{verbatim}
\usepackage{float}
\newcommand{\be}{\begin{equation}}
\newcommand{\ee}{\end{equation}}
\newcommand{\bea}{\begin{eqnarray}}
\newcommand{\eea}{\end{eqnarray}}
\def\4vol{{\int d^4x \sqrt{-g}}}
\def\lag{{\mathscr{L}}}

\def\heff{{\mathcal{H}_{eff}}}
\def\simlt{\stackrel{<}{{}_\sim}}
\def\simgt{\stackrel{>}{{}_\sim}}

\newcommand{\nc}{\newcommand} 
 
\nc{\lsim}{\begin{array}{c}\,\sim\vspace{-21pt}\\< \end{array}}
\nc{\gsim}{\begin{array}{c}\sim\vspace{-21pt}\\> \end{array}}

\nc{\LL}{L}
\nc{\vv}{\tilde{v}}
\nc{\GG}{\tilde{G}}

\title{
\vspace*{-1.3cm}
\begin{flushright}
\normalsize{
ANL-HEP-PR-03-057\\
EFI-03-39\\
  }
\end{flushright}
\vspace{1.5cm}
\Large
\textbf{ Beautiful Mirrors, Unification of 
Couplings \\
and Collider Phenomenology
}\vspace*{1.0cm}
\author{\large\textbf{D.E. Morrissey$^{a,b}$}, 
and \textbf{C.E.M.~Wagner$^{a,b}$}\\ \\[0.5cm]
$^a$\normalsize\emph{HEP Division, Argonne National Laboratory,
9700 Cass Ave.,
Argonne, IL 60439, USA} \\
$^b$\normalsize\emph{Enrico Fermi Institute, Univ. of Chicago, 5640
Ellis Ave., Chicago, IL 60637, USA}}}
\begin{document}
\setcounter{page}{0}
\maketitle
\begin{abstract}
The Standard Model provides an excellent description of the 
observables measured at high energy lepton and hadron colliders. However,
measurements of the forward-backward asymmetry of the bottom quark
at LEP suggest that the effective coupling of the 
right-handed bottom quark 
to the neutral weak gauge boson is significantly different from 
the value predicted by the Standard Model. Such a large 
discrepancy may be the result of a mixing of the bottom quark
with heavy mirror fermions with masses of the order of the weak
scale. To be consistent with the precision electroweak data,
the minimal extension of the Standard Model
requires the presence of vector-like pairs of 
$SU(2)$ doublet and singlet quarks. In this article, we show that
such an extension of the Standard Model is consistent with
the unification of gauge couplings and leads 
to a very rich phenomenology
at the Tevatron, the B-factories and the LHC. In particular,
if the Higgs boson mass lies in the range
120~GeV~$\simlt m_h \simlt 180$~GeV,
we show that 
Run II of the Tevatron collider with 
4--8~fb$^{-1}$ of integrated luminosity will have the potential
to discover the heavy quarks, while observing a 3-$\sigma$ evidence
of the Higgs boson in most of the parameter space.

\end{abstract}

\thispagestyle{empty}
\newpage

\setcounter{page}{1}

\baselineskip18pt

\section{Introduction}    

  In the absence of direct evidence for physics beyond the Standard 
  Model (SM), precision electroweak tests are the best way
  to get information about
  the scale and nature of a possible breakdown of the 
  SM description. While the SM has held firm in the face of a great   
  number of precision electroweak tests, the model has not emerged
  completely unscathed.  Fits of the SM to electroweak data show about a
  2.5-$\sigma$ deviation in the b-quark forward-backward asymmetry
  ($A^b_{FB}$)~\cite{EWWG}, and this situation has not improved much 
  in the last five  years.  This discrepancy is important for 
  two reasons.  On one hand, 
  it seems to indicate a significant deviation of the coupling of
  the right-handed bottom quark to the $Z$-gauge 
  boson (see, for example, Ref.~\cite{Haber:2000zh}). 
  On the other, this measurement plays an important role in the present 
  fits to the SM Higgs mass;
  the removal of the heavy quark data from
  the electroweak fits would push the central values of the Higgs mass
  to lower values, further inside the region excluded by the 
  LEP2 searches~\cite{Chanowitz:2001bv}.

  There are two ways of solving this apparent discrepancy, and both
  of them seem to indicate the presence of new physics. In 
  Ref.~\cite{Altarelli:2001wx}
  it was proposed to exclude the
  heavy quark data while introducing new physics that raises the
  central value of the Higgs boson mass, and improves the 
  fit to the other observables.
  Such a task requires new physics
  that gives a negative contribution to the $S$ parameter, positive
  contributions to the $U$ parameter and a moderate 
  contribution to the $T$ parameter. At least two examples of
  this kind of physics have been presented in the 
  literature~\cite{Altarelli:2001wx},~\cite{Carena:2002dz}; 
  the first within low energy 
  supersymmetry and the second within a warped extra-dimension scenario.

  An alternative to this procedure is to take seriously the heavy
  quark data while introducing new physics that modifies in a significant
  way the right-handed bottom quark coupling to the $Z$. 
  The Beautiful Mirror model of 
  Ref.~\cite{0a} accomplishes this by
  allowing the b-quark to mix strongly  with a set of exotic
  vector-like quarks.  This model turns out to have several other
  interesting features which we investigate in this paper.  To be
  specific, we consider the unification of gauge couplings, additional
  patterns of flavour mixing, the Higgs phenomenology, and searches
  for the heavy vector quarks.

  The model consists of the SM plus additional vector-like ``mirror''
  quarks.  These are a pair of $SU(2)$ doublets, $\Psi_{L,R} =
  {\chi_{L,R}' \choose \omega_{L,R}'}$, and a pair of $SU(2)$ singlets,
  $\xi_{L,R}'$.  Here and in what follows we use primed fields to
  denote gauge eigenstates, while mass eigenstates are written as
  unprimed fields.  The gauge group quantum numbers are the same as
  those of the analogous SM particles: $(3,2,1/6)$ for the doublets,
  and $(3,1,-1/3)$ for the singlets.  Since the quarks are added in
  vector-like pairs, these can have gauge-invariant Dirac masses,
  and the model is free of anomalies.  This is a minimal set of mirror
  quarks needed to improve the fit to electroweak data.

     The Yukawa and mass couplings of the mirror quarks are taken to be
\bea
                \label{coup}
        \lag &\supset& -(y_b\bar{Q}_{L}' + y_2\bar{\Psi}_{L}')b_{R}'\Phi 
        - (y_t\bar{Q}_{L}' + y_4\bar{\Psi}_{L}')t_{R}'\tilde{\Phi} 
        -  M_1\bar{\Psi}_{L}'\Psi_{R}' \\
        & & \phantom{aa} -(y_3\bar{Q}_{L}' + y_5\bar{\Psi}_{L}')\xi_{R}'\Phi 
        - M_2\bar{\xi}_{L}'\xi_{R}' + (h.c.)\nonumber
\eea
where $Q_{L}' = {t_{L}' \choose b_{L}'}$ is the usual third generation
SM quark doublet, and $\Phi = {\phi^+ \choose \phi^0}$ is the Higgs
doublet.  This is the most general set of renormalizable couplings
provided the mirror quarks couple only to each other and to the third
SM generation.\footnote{Note that couplings like
$\bar{Q}_{L}'\Psi_{R}'$ and $\bar{\xi}_{L}'b_{R}'$ can be rotated
away.}  As pointed out in~\cite{0a}, the Yukawa couplings $y_b$, $y_3$
and $y_4$ are constrained to be much smaller than $y_2$. Adjusting the ratio
$(\frac{v}{\sqrt{2}}y_2)/M_1 \simeq 0.7$, where $v = 246.22$~GeV is
the Higgs VEV, gives the best fit to precision electroweak data while
reducing the discrepancy in $A_{FB}^b$ to about one standard
deviation, 
and keeping the left-right b-quark asymmetry within one standard 
deviation of the value measured at SLC.  
This forces $y_2$ to be $\mathcal{O}(1)$ since $M_1
\gtrsim 200$~GeV is needed to explain why mirror quarks have not yet
been observed.  On the other hand, there are no strong constraints on
$y_5$, which mixes the exotics and therefore has only a small effect 
on the SM sector of the model. Following~\cite{0a}, we will mostly neglect 
$y_5$ for simplicity, altough we will comment on the effects of this 
coupling whenever they are relevant.

   This paper consists of seven sections.  
In Section~2 we examine the running of the gauge couplings 
and their unification at a high scale.  
In section~3 we discuss the issue of flavour mixing as well as
the quark couplings to the 
neutral and charged weak gauge bosons, and the Higgs.
Section~4 consists of an
investigation of the Higgs phenomenology in the model.  
In Section~5 we review 
the current limits on exotic quarks and investigate the possibility
of finding mirror quarks at the Tevatron.  
In Section~6 we examine how the new types of flavour mixing
possible with mirror quarks can affect CP
violation in $B \to \phi K_s$ decays.  
Finally, Section~7 
is reserved for our conclusions.

\section{Unification of Gauge Couplings}

The idea that the low energy gauge forces proceed from
a single grand unified description is a very attractive one,
and is supported by the apparent convergence of the weak,
hypercharge and strong couplings at short distances.  
The interest in low energy supersymmetry, for instance, 
has been greatly enhanced by
the discovery that the value of the strong coupling, $\alpha_s(M_Z)$,
can be deduced if one assumes that the gauge couplings unify at a high scale.  
This prediction depends on model-dependent
threshold corrections at the GUT scale, but to within the natural uncertainty
in these corrections~\cite{7a}, the predicted value of $\alpha_s(M_Z)$ is perfectly 
consistent with the values measured at low energies.
In the Standard Model, instead, the assumption of gauge coupling 
unification leads to a prediction for $\alpha_s(M_Z)$ that differs from
the measured value by an amount that is well
beyond the natural uncertainties induced by threshold corrections.

In~\cite{0a} it was noted that, to one-loop order, 
adding mirror quarks of the type considered here to the SM
greatly improves the prediction of $\alpha_s(M_Z)$ 
based on the assumption of gauge coupling unification. 
We extend this analysis by including the two-loop contributions to 
the gauge coupling
beta functions and the low-scale threshold corrections.  Since, for the
consistency of this study, the Higgs sector must remain weakly-coupled
while the Higgs potential should remain stable up to scales of the order
of the unification scale, $M_{G}$, we also
investigate the related issues of stability and perturbative
consistency of the Higgs sector.

In extrapolating the low energy description of the theory to 
short distances, it is important to remark that the Beautiful Mirror 
model~\cite{0a} does not provide a solution to the hierarchy
problem. Therefore, a main assumption behind this extrapolation is that
the physics that leads to an explanation of the hierarchy problem does not
affect the connection of the low energy couplings to the fundamental ones.
An example of such a theory construction is provided by warped extra 
dimensions~\cite{Randall:1999ee}, 
and has been investigated by several 
authors~\cite{gaugerunning}--\cite{Agashe:2002pr}.
In order to preserve the good agreement with the precision electroweak
data, the Kaluza-Klein modes must be heavier than a few TeV 
in this case~\cite{Carena:2003fx}, and 
therefore the low energy physics analyzed in the subsequent sections will
not be affected. On the other hand, extra dimensions could modify the
proton decay rate in a significant way by introducing new baryon number
violating operators, and, in the case of warped extra dimensions with
a Higgs field located in the infrared brane, would make the issue of
the running of the Higgs quartic coupling an irrelevant one. For the
rest of this section, we shall proceed with a pure four dimensional
analysis of the evolution of couplings and of the proton decay rate.

\subsection{Renormalization Group Equations}

  Using the results of~\cite{1a,2a}, the two-loop ($\overline{MS}$ scheme)
gauge coupling beta functions are
\be 
\beta_l =
\frac{dg_l}{dt} = -\frac{1}{(4\pi)^2}b_lg_{l}^{3} -
\frac{1}{(4\pi)^4}\sum_{k=1}^3b_{kl}g_{k}^2g_{l}^2 -
\frac{1}{(4\pi)^4}g_{l}^3Y_4^l(F)
\ee 
where $t =
\ln\left(\frac{\mu}{M_Z}\right)$ is the energy scale, and $l = 1, 2,
3$ refers to the U(1), SU(2), and SU(3) gauge groups respectively.
The first term is the one-loop contribution, while the other terms
come from two-loop corrections.

The coefficients $b_l$ and $b_{kl}$ are given by
\be
        \begin{array}{ccc} b_1 = -\frac{9}{2},&b_2 = \frac{7}{6},&
b_3 = 5,\end{array}\nonumber
\ee
and
\be
        b_{kl} = -\left(\begin{array}{ccc}
                        \frac{291}{25}&1&\frac{13}{10}\\
                        3&\frac{91}{3}&\frac{15}{2}\\
                        \frac{52}{5}&20&12
                        \end{array} \right).\nonumber
\ee
In the SM, the corresponding one-loop beta function coefficients
are $b_1^{SM} = -41/10$, $b_2^{SM} = 19/6$ and $b_3^{SM} = 7$. The
variation of these coefficients are hence $\Delta b_1 = 2/5$ and
$\Delta b_2 = \Delta b_3 = 2$. Since $b_2$ and $b_3$ are shifted by
an equal amount, they tend to unify at the same scale 
as in the SM, about a few times $10^{16}$~GeV. Interestingly enough,
the shift in $b_1$ is much smaller than that of $b_2$ and $b_3$,
leading to, as we shall see, a successful unification of the three couplings.

The coefficients $Y_4^l(F)$ involve the Yukawa couplings.  
Neglecting the small Yukawa couplings $y_b, y_3, y_4,$ and $y_5$, they are 
\be
        Y_4^l(F) = C_{lt}{y_t}^2 + C_{l2}{y_2}^2 
\ee
where 
\be
        C_{lf} = \left(\begin{array}{cc}
                        \frac{17}{10}&\frac{1}{2}\\
                        \frac{3}{2}&\frac{3}{2}\\
                        2&2
                        \end{array} \right)\nonumber
        \label{clf}
\ee
and f = t,2.

  The Yukawa couplings evolve according to
\be
        (4\pi)^2\frac{dy_f}{dt} = \beta_fy_f
\ee
where $f = t,2$.  The one-loop and leading two-loop contributions to
$\beta_f$ were calculated following~$\cite{3a}$.  Of the two-loop
terms, we include only those involving $g_3$ or the Higgs self
coupling $\lambda$; the $g_3$ terms are enhanced by large colour
factors while the $\lambda$ terms can become important when
investigating the stability of this coupling.  The one-loop
contributions are
\bea
        (4\pi)^2\beta_t^{(1)}  &=&  
        \frac{9}{2}y_t^2 + 3y_2^2 - 
        \left(\frac{17}{20}g_1^2 + \frac{9}{4}g_2^2 
        + 8g_3^2\right), \nonumber\\
        (4\pi)^2\beta_2^{(1)}  &=&  
        \frac{9}{2}y_2^2 + 3y_t^2 - \left(\frac{1}{4}g_1^2 
        + \frac{9}{4}g_2^2 + 8g_3^2\right).
\eea
The two-loop contributions that we have included are
\bea
        (4\pi)^4\beta_t^{(2)}  
        &=&  \frac{3}{2}\lambda^2 
        - 6y_t^2\lambda + g_3^2(46y_t^2 + 20y_2^2)  
        - \frac{284}{3}g_3^4, \nonumber\\
        (4\pi)^4\beta_2^{(2)}  
        &=&  \frac{3}{2}\lambda^2 - 6y_2^2\lambda 
        + g_3^2(20y_t^2 + 46y_2^2)  - \frac{284}{3}g_3^4. 
\eea
The total beta-function is the sum of these pieces: $\beta_f =
\beta_f^{(1)} + \beta_f^{(2)}$.  Aside from the modifications due to
the mirror quarks, these are in agreement with the results 
of~\cite{1a}.

   The Higgs self-coupling $\lambda$ is taken to be \be \lag \supset
\mu^2 \Phi^{\dag}\Phi- \frac{1}{2}\lambda(\Phi^{\dag}\Phi)^2.  \ee 

With this definition, the tree-level Higgs mass is $m_h =
\sqrt{\lambda}v$,
 where $v = 246.22~\textrm{GeV} = \sqrt{2}\left<\phi^0\right>$.  
$\lambda$ evolves according to $\frac{d\lambda}{dt} = \beta_{\lambda}$.
We have calculated the one-loop and leading two-loop 
contributions to $\beta_{\lambda}$ using the results of~\cite{4a, 5a}.  
As for the Yukawas, only the largest two-loop terms 
involving $g_3$ or $\lambda$ were included.  
For the one-loop part, we obtain
\bea
        (4\pi)^2\beta_{\lambda}^{(1)} &=&  
        12\lambda^2 - \left(\frac{9}{5}g_1^2 + 9g_2^2\right)\lambda 
        + \frac{9}{4}\left(\frac{3}{25}g_1^4 
        + \frac{2}{5}g_1^2g_2^2 + g_2^4\right) \nonumber\\
        & &     {}+ 12\lambda \left(y_t^2 + y_2^2 \right) 
        - 12\left( y_t^4 + y_2^4 \right).
\eea
The two loop part is given by
\bea
        (4\pi)^4\beta_{\lambda}^{(2)} &=& -78\lambda^3 - 72(y_t^2 + y_2^2)\lambda^2 - 3(y_t^4 + y_2^4)\lambda + 60(y_t^6 + y_2^6)\nonumber\\
                                & & \phantom{a}  + 18(\frac{3}{5}g_1^2 + 3g_2^2)\lambda^2  + 80 g_3^2(y_t^2 + y_2^2)\lambda - 64g_3^2(y_t^4 + y_2^4).
\eea
Again, the total beta-function is the sum of the one- and two-loop parts.

\subsection{Input Parameters and Threshold Corrections}

   We have investigated the running of these couplings numerically.
The initial values $\overline{MS}$ scheme values were
taken from~\cite{6a}:
\be
        \begin{array}{cclccl}
        \alpha^{-1}(M_Z) &=& 127.922 \pm 0.027&
        \phantom{aaa}\sin^2\theta_w(M_Z) &=& 0.23113 \pm 0.00015\\
        M_Z &=& 91.1876 \pm 0.0021 &\phantom{aaa}v&=&
        246.22\phantom{a}\textrm{GeV} \\
        \bar{m}_t(m_t) &=& 165 \pm 5 
        \phantom{a}\textrm{GeV}.&&\nonumber
        \end{array}
\ee
These parameters correspond to the effective $SU(3)_c\times U(1)_{em}$ theory
with five quarks obtained by integrating out the heavy gauge bosons
and quarks in the full $SU(3)_c\times SU(2)\times U(1)_Y$ theory at scale
$M_Z$.  Threshold corrections to the gauge couplings arise in the
process of matching the theories.  We define
\bea
        \tilde{\alpha}_1^{-1} &=& \frac{3}{5}(1 - \sin^2\theta_w)\alpha^{-1},\\
        \tilde{\alpha}_2^{-1} &=& \sin^2\theta_w\alpha^{-1}\nonumber,\\
        \tilde{\alpha}_3^{-1} &=& \alpha_s^{-1}\nonumber.
\eea
Then the gauge couplings at $M_Z$ are given by
\be
        \alpha_l^{-1} = \tilde{\alpha}_l^{-1} + \rho_l,
\ee
where the $\rho_l$ terms represent threshold corrections.  
To one-loop order, these are~\cite{7a} 
\bea
        \rho_3 &=& 
        \frac{1}{3\pi}\sum_\zeta \ln(\frac{m_{\zeta}}{M_z}),\nonumber\\
        \rho_2 &=& 
        \sin^2\theta_w\left[\frac{1}{6\pi}(1 - 21 \ln(\frac{M_W}{M_Z})) 
        + \frac{2}{\pi}\sum_\zeta Q_{\zeta}^2 
        \ln(\frac{m_{\zeta}}{M_Z})\right],\nonumber\\
        \rho_1 &=& 
        \frac{3}{5}\left(\frac{1 - \sin^2\theta_w}{\sin^2\theta_w}\right)\rho_2,
\eea
where the sums run over $\zeta = t, \chi, \omega, \xi$.  As shown in~\cite{0a}, the tree-level masses of the mirror quarks are given by
\be
        \begin{array}{lclrcl}
        m_{\chi} &=& M_1,&\phantom{aa}m_{\omega} &=& \left(M_1^2 + Y_2^2\right)^{1/2},\\
        m_{\xi} &=& M_2,&&&
        \end{array}
\label{mass}
\ee
where $Y_2 = \frac{v}{\sqrt{2}}y_2$.  These parameters are not
completely independent.  As explained above,
$Y_2 \simeq 0.7 \; M_1$~\cite{0a} gives the best fit to
electroweak data, while $M_2, M_1 \gtrsim 200$~GeV are needed to
explain why these exotics have not yet been observed at the
Tevatron~\cite{affolder} (see section 5).

\begin{figure}
\centerline{
        \includegraphics[width=0.7\textwidth]{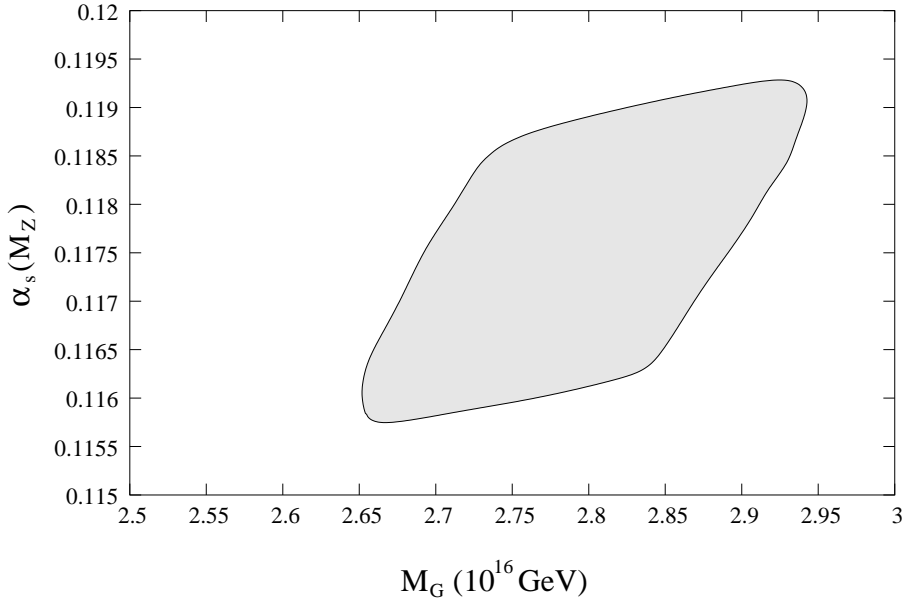}}
        \caption{Range of $\alpha_s$ consistent with a 1$\%$ unification 
        of the gauge couplings plotted against the unification scale $M_G$.
        \label{unif}}
        
\end{figure}

\subsection{Numerical Evolution}

   The unification of gauge couplings was investigated by fixing
   $\sin^2\theta_w(M_Z)$ and  $\alpha_{em}(M_Z)$ according to their measured
   values, and varying $\alpha_s(M_Z)$ until the gauge couplings
   unified to within $1\%$.  GUT-scale threshold corrections were not
   considered.  Figure \ref{unif} shows the range of $\alpha_s(M_Z)$
   obtained in this way for 250 GeV $\leq M_2 \leq$ 1000 GeV and all
   values of $\lambda(M_Z)$ and $y_2(M_Z)$ consistent with unification.
   (See the following section.)  The range is
   plotted against the unification scale.  In general, the unification
   is quite insensitive to the input values of $M_2, \lambda$, and
   $y_2$.  The scale of unification is quite high, $M_G = (2.80 \pm
   0.15)\times 10^{16}$~GeV, depending on the input values, at which
   point the unified gauge coupling constant has value $\alpha_G^{-1}
   = 35.11 \pm 0.05$. 

The predicted range of the strong gauge coupling  
is in excellent agreement with the
values measured experimentally. This agreement is quite intriguing 
since the particular completion of the Standard Model
considered in this work
is motivated by data and not by any model building consideration. 
We shall not attempt to construct a grand unified model leading to
the appearance of the  mirror quarks considered here in 
the low energy spectrum. 
Instead, we will concentrate on additional
issues regarding the renormalization group evolution of the dimensionless
couplings of the theory, as well as on exploring general features of the
low energy phenomenology of this particular extension of the Standard
Model.   

\subsection{Stability and Non-Triviality of the Higgs}

\begin{figure}[!htb]
\centerline{
        \includegraphics[width=0.7\textwidth]{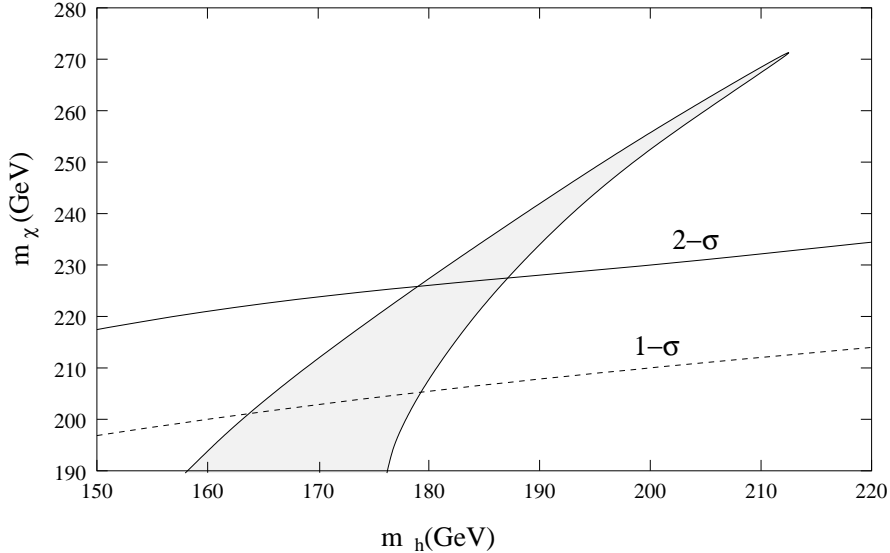}}
        \caption{Region in the $m_h$-$m_{\chi}$ plane that 
is consistent with unification (shaded region) and precision
electroweak data (below the dashed and solid lines).
        \label{pspace}}
\end{figure}

   If the extrapolation of the model up to high scales is to be
   self-consistent, it should remain stable and weakly-coupled up to
   the unification scale.  The only source of trouble in this 
   regard is the Higgs self-coupling $\lambda$.  
   Stability of the Higgs sector requires $\lambda(Q) > 0$, 
   for all $Q < M_{G}$ while perturbative
   consistency means that $\lambda$ must not be too large.  For
   concreteness we demand~\footnote{This upper 
   limit on $\lambda$ is somewhat arbitrary
   but fairly unimportant in the present case since $\lambda$ grows
   very quickly when it becomes larger than unity.} 
   that $0 < \lambda < 2$ up to $10^{17}$~GeV.
   This is
   sufficient to guarantee that the effective potential is similarly
   well behaved~\cite{framp}.  The evolution of $\lambda$ is largely
   determined by the initial values of $y_2$ and $\lambda$.  Only for
   a small subset of initial values does $\lambda$ remain well-behaved
   (i.e. $0<\lambda<2$) up to $M_G$.  This subset is shown in Figure
   \ref{pspace}, where we have written $\lambda$ in terms of the
   (tree-level) Higgs mass, and $Y_2$ in terms of $m_{\chi}$
   assuming $Y_2 = 0.7M_1$~\cite{0a}.

We compare the allowed region with the region favoured by precision
electroweak data found in~\cite{0a}.  There is a small overlap between
the allowed band found here and the 1-$\sigma$ allowed region of~\cite{0a} 
corresponding to $160~\mbox{GeV} \lesssim m_h \lesssim 180$ GeV 
and $m_{\chi} \lesssim225$~GeV ($m_{\omega} \lesssim 275$~GeV).  
We show in sections~4 and 5 that these mass ranges will
be probed by Run~II at the Tevatron with 
$4-8~\mbox{fb}^{-1}$ of data. 

Finally, we note that including
the $y_5$ coupling (see Eq.~(\ref{coup})) would tend to displace the shaded
region in Figure~\ref{pspace} downwards, slightly increasing the
preferred range of Higgs masses.  This is because, to one loop order,
 the beta functions are modified as $y_2^2 \to (y_2^2+y_5^2)$
by this inclusion, while the best fit value of $Y_2/M_1$ changes very little
(see section~\ref{mqmass}). Thus, our bound on $y_2$ obtained with $y_5 = 0$ 
translates into a bound on $\sqrt{y_2^2+y_5^2}$ in the more general case,
leading to lower values of $m_{\chi}$ for a given $m_h$.

\subsection{Proton Decay}

  Grand unified models induce baryon number %
violating operators that lead
to a proton decay rate that may be observable in the next generation
of proton decay experiments. The present bounds on the proton 
lifetime~\cite{superK} already put relevant bounds on grand 
unified scenarios. 
In four dimensional supersymmetric grand unified models, for instance,  
dimension five operators may easily induce a proton
lifetime shorter than the present experimental bound~\cite{ProtDecay}.
This situation may be avoided 
by a suitable choice of the low
energy spectrum~\cite{raby}. Heavy first and second generation 
sfermions and light gauginos are preferred from these considerations. 
 
In the model under study there are no dimension-five operators, 
so the dominant decay mode is expected to be $p \to \pi^0e^+$. 
The high unification scale obtained
above means that the proton will be long-lived regardless of the
details of the unification mechanism.  If proton decay proceeds via $SU(5)$
gauge bosons, the decay rate is given by~\cite{emcosta}
\be
        \Gamma(p\to \pi^0e^+) = 
        \frac{\pi m_p \alpha_G^2}{8f_{\pi}^2M_G^4}(1 + D + F)^2\alpha_N^2
        \left[ A_R^2  + (1+|V_{ud}|^2)^2A_L^2\right]
\ee
where $f_{\pi} = 0.131$ GeV is the pion decay constant, $D = 0.81$ and
$F =0.44$ are chiral Lagrangian factors, $\alpha_N$ is a
coefficient related to the $\pi^0 p$ operator matrix element, and
$A_L$ and $A_R$ are correction factors due to the running of the
couplings.  A recent lattice-QCD calculation gives~\cite{aoki}
$|\alpha_N| = 0.015(1)$ GeV$^3$, where the
uncertainty is purely statistical.  The systematic uncertainty is
probably much larger; we take it to be $\sim 50\%$~\cite{raby}. The
correction factors $A^{L,R}$ 
split into long and short distance
pieces: $A_{L,R} = A_l\prod_{i=1}^3A_i^{L,R}$, where $A_l$ comes from the
renormalization group evolution below $M_Z$ and $A_i^{L,R}$ from that
above.  Here, $A_l \simeq 1.3$ is identical to the SM value, while the
short distance pieces, to one-loop order, are~\cite{wilc} 
\be
        \begin{array}{cclcccc}
        A_3^{L} &=&  
\left[\frac{\alpha_3(M_Z)}{\alpha_G}\right]^{6/(33-4n_g-6)} 
&\simeq& 3.15 &=&A_3^{R},\nonumber\\
        A_2^{L} &=& \left[\frac{\alpha_2(M_Z)}{\alpha_G}\right]^{27/(86-16n_g-24)} 
&\simeq& 1.39 &=& A_2^{R},\nonumber\\
        A_1^{L} &=& 
\left[\frac{\alpha_1(M_Z)}{\alpha_G}\right]^{-69/(6+80n_g+24)} &\simeq& 1.14,\nonumber\\
        A_1^{R} &=& 
\left[\frac{\alpha_1(M_Z)}{\alpha_G}\right]^{-33/(6+80n_g+24)} &\simeq& 1.07.
        \end{array}
\ee

Using $M_G = 2.8\times 10^{16}$ GeV, and $\alpha_G^{-1} = 35.1$, we find
\be
        \tau(p\to\pi^0e^+) = 3\times10^{36\pm1} {\rm yrs},
\ee     
well in excess of the Super-Kamiokande bound of 
$\tau(p\to\pi^0e^+) = 5.3\times10^{33}$ yrs~\cite{superK}.

\section{Flavour Mixing}

  Extending the matter content of the SM also introduces new sources
  of flavour mixing.  With mirror quarks, this pattern can be quite
  complicated, involving right-handed couplings to the W, and
  tree-level flavour-changing couplings to the Z and the Higgs.  We
  consider first the generic case, taking the most general set of
  Yukawa and mass terms possible.  Next, we simplify our results
  making use of the fact that, in the model under study, 
  the mirror quarks couple only to the third generation quarks
  and calculate explicitly the couplings of the heavy quarks to the 
  weak gauge bosons and the Higgs.  
  In subsequent sections we shall use these results to investigate the Higgs 
  phenomenology, the collider searches for mirror quarks, and CP
  violation in $B \to \phi K_s$ decays.

  Let $\lambda_u$ and $\lambda_d$ be the flavour-space mass matrices
describing the flavour mixing between gauge eigenstates. These matrices
will be $4\times 4$ and $5\times 5$ respectively, and will have
contributions from Yukawa couplings and Dirac mass terms.  Both
matrices can be diagonalized by bi-unitary transformations: 
\be
        \begin{array}{cc}
        \lambda_u = U_uD_uW^{\dagger}_u,&\lambda_d = U_dD_dW^{\dagger}_d,
        \end{array}
\ee
where the $U$'s and $W$'s are unitary, and $D_u$ and $D_d$ are the
diagonalized mass matrices.  The corresponding (unprimed) mass
eigenstates are then related to the (primed) gauge eigenstates by the
unitary transformations
\be
        \begin{array}{cc}
        {u'}^A_L = U_u^{AB}u_L^B,&\phantom{a}{u'}^A_R = W_u^{AB}u_R^B,\\
        {d'}^P_L = U_d^{PQ}d_L^Q,&\phantom{a}{d'}^P_R = W_d^{PQ}d_R^Q.
        \end{array}
\ee
Here, the indices $A,B = 1,\ldots 4$ correspond to $\{u,c,t,\chi\}$,
while $P,Q = 1,\ldots 5$ refer to $\{d,s,b,\omega,\xi\}$ respectively.

In terms of the physical (mass eigenstate) fields, the charged
currents become 
\bea \sqrt{2}J_{W^+}^{\mu} &=&
\bar{u}^{B}_L\gamma^{\mu}V_{L}^{BQ}d^{Q}_{L} +
\bar{u}^{B}_{R}\gamma^{\mu}V_{R}^{BQ}d^{Q}_{R},\nonumber\\ 
J_{W^-}^{\mu} &=&
{J_{W^+}^{\mu}}^{\dagger}, \eea 
where the $4\times 5$
flavour-mixing matrices are given by 
\bea V_{L}^{BQ} &=&
\sum_{i=1}^4{U_{u}^{iB}}^*U_{d}^{iQ},\nonumber\\ 
V_{R}^{BQ} &=&
{W_{u}^{B4}}^*W_{d}^{4Q}. 
\eea 
The matrix $V_L$ is analogous
to the CKM matrix $V_{CKM}$ of the SM.  It is nearly unitary in the
sense that $V_L{V_L}^{\dagger} = \mathbb{I}_{4\times 4}$ and
$V_L^{\dagger}V_L = \mathbb{I}_{5\times 5} - V_d$, where the matrix
$V_d$ is defined below.  The matrix $V_R$ describes right-handed
couplings, and has no analogue in the SM.

Similarly, the hadronic neutral current is
\bea
\cos\theta_wJ^{\mu}_Z &=& 
\bar{u}^A_L\gamma^{\mu}(\frac{1}{2} - \frac{2}{3}\sin^2\theta_w)u_L^A +
\bar{u}^A_R\gamma^{\mu}(-\frac{2}{3}\sin^2\theta_w)u_R^A\nonumber\\             
& & + \bar{d}^P_L\gamma^{\mu}(-\frac{1}{2} 
+ \frac{1}{3}\sin^2\theta_w)d_L^P + \bar{d}^P_R\gamma^{\mu}
(\frac{1}{3}\sin^2\theta_w)d_R^A\nonumber\\
& & \phantom{aa} + \frac{1}{2} \left(
\bar{u}^A_R\gamma^{\mu}V_u^{AB}u_R^B
 - \bar{d}^P_R\gamma^{\mu}\tilde{V}^{PQ}_dd^Q_R +
\bar{d}^P_L\gamma^{\mu}V^{PQ}_dd^Q_L \right) 
\eea
where the matrices $V_u, V_d, \tilde{V}_d$ are given by
\bea
        V_u^{AB} &=& {{W}^{4A}_u}^*W^{4B}_u,\\
        V_d^{PQ} &=& {{U}^{5P}_d}^*U^{5Q}_d,\nonumber\\
        \tilde{V}_d^{PQ} &=& {{W}^{4P}_d}^*W^{4Q}_d.\nonumber
\eea
The off-diagonal elements of these matrices describe FCNCs.  Each is
Hermitian and satisfies $V^2 = V$.

\subsection{Heavy Quark Neutral and Charged Currents}

The expressions above can be simplified considerably by
using the fact that, 
in the model under study,
the (gauge eigenstate) mirror quarks 
couple only to the quarks of the third generation  
(see Eq.~(\ref{coup})).  The mixing between the 
mirror quarks and the first and second generation quarks
is thus very small, and so will be neglected.  
Moreover, the mixing
between the SM quarks is given approximately by the usual
CKM description.  The flavour violating effects among
the heavy quarks are then related to the mixing of 
the right-handed $SU(2)$ quark-doublet
with the third generation right-handed quarks, as well as
the mixing of the left-handed $SU(2)$ quark-singlet with the
left-handed bottom-quark. 

The mixing in the top sector must be very small
to avoid a conflict with the $\mathcal{B}(b \to s \gamma)$
predictions~\cite{Larios:1999au}. 
We shall therefore assume, for simplicity, 
vanishing mixing in the top sector ($y_4 =0$).  The top-sector
mass matrix is then diagonal, with $m_t = \frac{v}{\sqrt{2}}y_t$
and $m_{\chi} = M_1$.

Following~\cite{0a} we take $y_5 = 0$ as well.    
The bottom-sector mass matrix, in the basis $(b',\omega',\xi')$,
is then given by
\be
        \lambda_d =
        \left(\begin{array}{ccc}
        Y_b&0&Y_3\\
        Y_2&M_1&0\\
        0&0&M_2\end{array}
        \right)
\ee
where $Y_i = \frac{v}{\sqrt{2}}y_i$, $i = b,2,3$.  
The phenomenologically interesting regime is
$Y_b, Y_3 \ll Y_2, M_1, M_2$~\cite{0a}.  Working to linear order
in the small quantities $Y_b/M_1$ and $Y_3/M_2$, the 
left- and right-handed mixing matrices are 
\be
        U_d = \left(\begin{array}{ccc}
        c_L\tilde{c}_L&s_L&c_L\tilde{s}_L\\
        -s_L&c_L&0\\
        -\tilde{s}_L&0&\tilde{c}_L\end{array}\right)
\ee
and
\be
        W_d = \left(\begin{array}{ccc}
        \phantom{-}c_R&s_R&0\\
        -s_R&c_R&0\\
        \phantom{-}0&0&1\end{array}\right)      
\ee
where $s_R = \sin\theta_R$, $s_L = \sin\theta_L$, and 
$\tilde{s}_L = \sin\tilde{\theta}_L$ are given by
\bea
        s_R &=& \frac{Y_2}{(Y_2^2 + M_1^2)^{1/2}},\nonumber\\
        s_L &=& \frac{Y_bY_2}{(M_1^2 + Y_2^2)},\nonumber\\      
        \tilde{s}_L &=& \frac{Y_3}{M_2}.
\eea
Applying the mixing matrices to $\lambda_d$,
the b-sector masses are
\be
\begin{array}{cccccc}
        m_b &=& Y_b\left(1 + \frac{Y_2^2}{M_1^2}\right)^{-1/2},&&\\
        m_{\omega} &=& (M_1^2 + Y_2^2)^{1/2},&\phantom{a}m_{\xi} &=& M_2.       
\end{array}
\ee
To obtain a good agreement between the predictions of the model and 
precision electroweak data
the angle in the right-handed sector must be sizeable,
\be
        \tan\theta_R = Y_2/M_1 \simeq 0.7,
\label{tan}
\ee
while $\tilde{s}_L$ should be small,
\begin{equation}
\sin\tilde{\theta}_L \simeq 0.09.
\end{equation}
Note that Eq.~(\ref{tan}) fixes $s_L$ in terms of the b mass.

In this approximation, the relevant neutral currents read
\begin{eqnarray}
\cos\theta_wJ_{Z}^{\mu}& = & 
\bar{b}_R \gamma^{\mu} b_R \left(-\frac{s_R^2}{2} + 
\frac{\sin^2\theta_W}{3} \right) 
\nonumber\\
& + &
\bar{\omega}_R \gamma^{\mu} \omega_R \left(-\frac{c_R^2}{2} + 
\frac{\sin^2\theta_W}{3} \right) -
\left(\bar{b}_R \gamma^{\mu} \omega_R + h.c. \right) \frac{s_R c_R}{2}  
\nonumber\\
&+&
\bar{b}_L \gamma^{\mu} b_L \left(-\frac{\tilde{c}^2_R}{2} + 
\frac{\sin^2\theta_W}{3} \right) 
\nonumber\\
& + &
\bar{\xi}_L \gamma^{\mu} \xi_L \left(-\frac{\tilde{s}_L^2}{2} + 
\frac{\sin^2\theta_W}{3} \right) +
\left(\bar{b}_L \gamma^{\mu} \xi_L + h.c. \right) 
\frac{\tilde{s}_L\tilde{c}_L}{2}  
\nonumber\\
&+&
\bar{\omega}_L \gamma^{\mu} \omega_L \left(-\frac{1}{2} + 
\frac{\sin^2\theta_W}{3} \right) +
\bar{\xi}_R \gamma^{\mu} \xi_R  
\frac{\sin^2\theta_W}{3} .
\label{Jz}
\end{eqnarray}
Within the same approximation,
the charged currents read
\bea
J_{W^+}^{\mu} & = &
\bar{t}_L \gamma^{\mu} \left( \tilde{c}_L b_L + s_L \omega_L 
+ \tilde{s}_L \xi_L \right)
\nonumber\\
& + &
\bar{\chi}_L \gamma^{\mu} \left( \omega_L  - s_L  b_L \right)
\nonumber\\
& + &
\bar{\chi}_R \gamma^{\mu} \left( c_R \omega_R - s_R b_R \right),
\label{Jw}
\eea
where we have neglected terms of order $m_b^2/M_1^2$.

In the above we have neglected the $y_5$ effects. We have verified 
that even large values of $y_5$ produce a negligible
effect in the relevant right-handed $Zbb$ coupling, while inducing 
only a very small change in the left-handed $Zbb$ coupling
(that can be compensated by a minor change of $y_3$),
and so $Y_2/M_1 = 0.7$ still gives the bit fit to electroweak data.  
On the other hand, since $y_5$ induces a mixing between the two heavy mirror quarks,
the $Z$ couplings and the mass spectrum of the mirror quarks
are modified by this coupling. In particular, for non-vanishing
values of $y_5$ there is always
a mirror quark state in the b-sector with mass smaller than
$m_{\omega}^0 = (M_1^2 + Y_2^2)^{1/2}$.  Therefore, for any given
value of $m_{\chi}$, $m_{\omega}^0$~provides an upper bound on
the lightest mass of the down-type mirror quarks. With $Y_2/M_1 = 0.7$ 
and $M_1 < 250$~GeV as required by precision electroweak data~\cite{0a}, 
the lightest down-type mirror quark can not be heavier than 300~GeV.
\label{mqmass}

\subsection{Higgs Couplings}

    One of the most important immediate goals of high energy physics
is to understand the mechanism of electroweak symmetry breaking. 
In the Standard Model and its supersymmetric extensions,
this symmetry is broken spontaneously through the vacuum expectation
value of one or more scalar Higgs bosons. The same is true for the
model under study and it is therefore quite relevant to understand the
possible modification of the Higgs boson search channels at the Tevatron
and the LHC.

In addition to introducing new sources of flavour mixing, 
mirror quarks also modify the couplings to the Higgs.  
The Dirac mass terms for the mirror quarks mean that the Higgs-quark 
couplings need no longer be flavour diagonal in the basis of mass eigenstates, 
nor be proportional to the quark masses.  
We find that the coupling of the Higgs to the b quark is reduced relative to 
the SM.  This, along with the contribution of the heavy quarks in loops, 
has interesting consequences for the detection of the Higgs.  

As in the previous section, we will assume that the mirror quarks mix almost 
exclusively with the third generation quarks and take 
$y_4, y_5 \approx 0$.  
This implies that the only Higgs-quark coupling that is significantly 
modified from the SM is that of the b-quark.    
The relevant terms in the Lagrangian are therefore  
\be
\lag \supset -(y_b\bar{Q}_{L}' + 
y_2\bar{\Psi}_{L}')b_{R}'\Phi  - 
y_3 \bar{Q}_{L}' \xi_{R}' \Phi -
M_1\bar{\Psi}_{L}'\Psi_{R}' -
M_1\bar{\xi}_{L}'\xi_{R}'
+ (h.c.).
\ee
After symmetry breaking 
$\Phi =  \frac{1}{\sqrt{2}}{0 \choose v + h}$ in the unitary gauge.  These couplings 
can then be written as 
\be
        \lag \supset -\bar{d}_{L}'\left(M_b + 
        h N_b\right)d_{R}' + (h.c.)
\ee
with $d_{L,R}' = \left( b_{L,R}', \omega_{L,R}', 
\xi_{L,R}'\right)^t$, and  
\be
        N_b = \left(\begin{array}{ccc}
                        Y_b&0&Y_3\\
                        Y_2&0&0\\
                        0&0&0
         \end{array}\right).
\ee
Transforming to the mass eigenstate basis, the Higgs couplings become
\bea
\label{home}
        \lag &\supset& - c_R^2 \frac{m_b}{v}h\bar{b}_Lb_R - 
s_R^2 \frac{m_{\omega}}{v}h\bar{\omega}_L\omega_R \\
                & &  \phantom{a} - s_R c_R h\frac{m_b}{v}
\bar{b}_L\omega_R  - s_R c_R h\frac{m_{\omega}}{v}\bar{\omega}_Lb_R 
\nonumber\\
        & & \phantom{aa} -
\frac{m_\xi \tilde{s}_L}{v} h \left( \bar{b}_L +
        \tilde{s}_L \bar{\xi}_L \right) \xi_R \; + \; (h.c.).
\nonumber
\eea
Using the value $\tan\theta_R = Y_2/M_1 \simeq 0.7$ favoured by the model, we find that
the hbb coupling is reduced by a factor of $c_R^2 \sim 2/3$.

\section{Higgs Phenomenology}
\subsection{Higgs Production and Decay}

   This scenario modifies the phenomenology of the Higgs in two ways.
First, by reducing the hbb coupling by a factor of $c_R^2$, 
the partial width $\Gamma(h \to \bar{b}b)$ is attenuated
by the square of this factor, $c_R^4 \sim 4/9$.
Since this channel is dominant for Higgs masses below $m_h \simeq 130$~GeV,
the reduction of the partial width for this mode
increases the branching fractions of the other modes in this range.
Secondly, $\omega$ quark loop effects 
increase the partial width $\Gamma(h \to gg)$. 
 This increases both the branching fraction of this mode, 
and the Higgs production cross-section by gluon fusion.

\begin{figure}[hbt!]
\centerline{
\includegraphics[angle = 0, width = 0.7\textwidth]{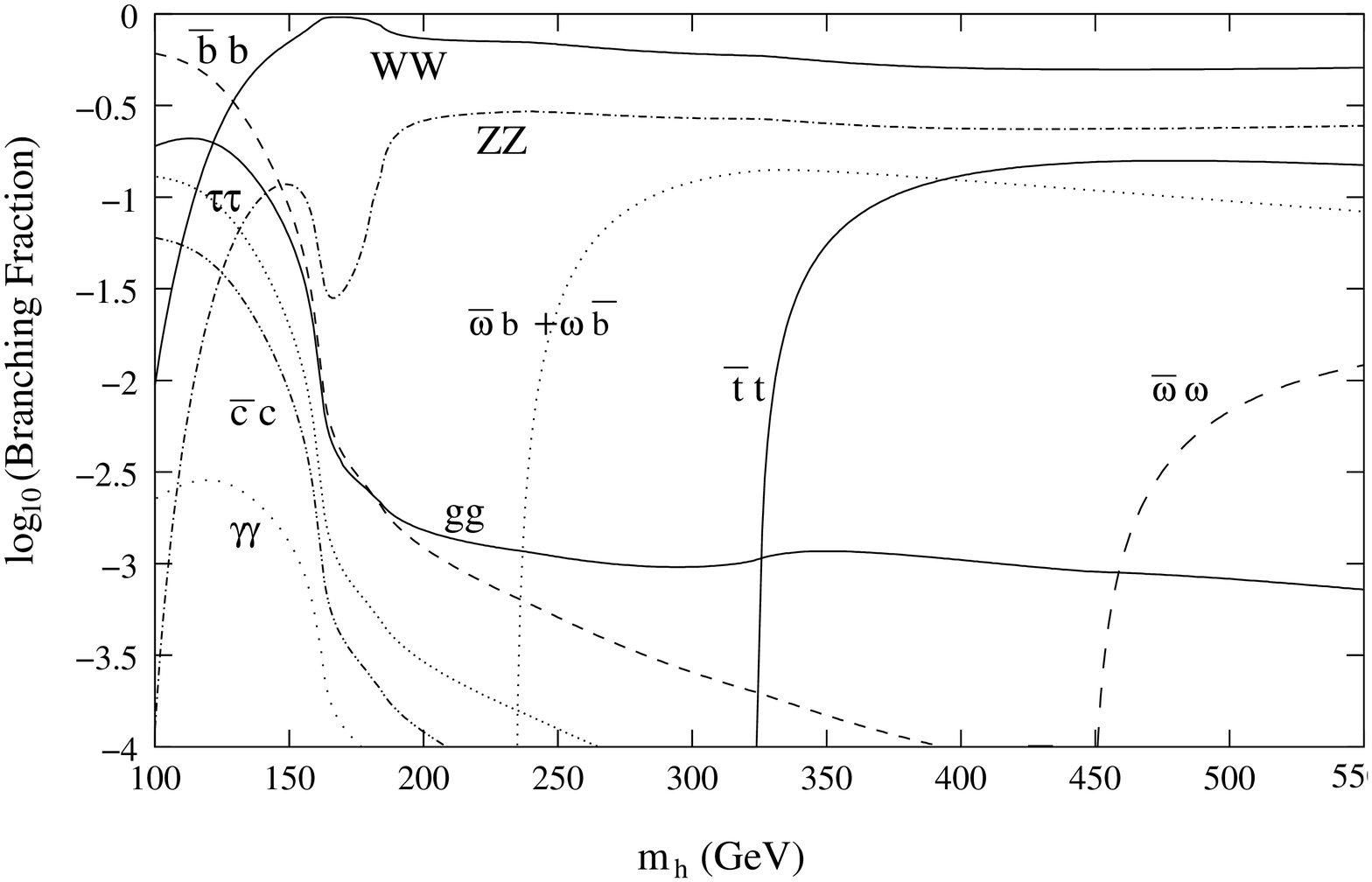}}
        \caption{Higgs Branching Ratios with Mirror Quarks,
        for $m_{\omega} = 250$ GeV and $Y_2/M_1 = 0.7$.
\label{brmq}}
\end{figure}

  Let us consider the effect of the $\omega$ quark in a bit more
detail.  The presence of this quark in a loop connecting the Higgs to
two gluons modifies the $h \to gg$ partial width.  Neglecting light
quark contributions, and keeping only the effects of the
dominant Yukawa coupling $y_2$, the partial width becomes 
\be 
\Gamma(h \to gg) =
\frac{\alpha\alpha_s^2}{128\pi^2\sin^2\theta_w}
\left(\frac{m_h^3}{M_W^2}\right)
\left|F_{1/2}(\tau_t)
+ s_R^2F_{1/2}(\tau_{\omega})\right|^2
\label{hloop} 
\ee 
where
the function $F_{1/2}(\tau_q)$ is given by~\cite{hhunt} 
\be 
F_{1/2} =
-2\tau_q[1 + (1 - \tau_q)f(\tau)], 
\ee 
with $\tau_q =
4(\frac{m_q}{m_h})^2$ and 
\be f(\tau) = \left\{ \begin{array}{ccc}
[\sin^{-1}(1/\sqrt{\tau})]^2;&\tau \ge 1&\\
-\frac{1}{4}[\ln(\eta_+/\eta_-) - i\pi]^2;&\tau < 1;& \eta_{\pm} = 1
\pm \sqrt{1 - \tau}.
                \end{array} \right.  
\ee
The corresponding expression in the SM is obtained by setting $s_R = 0$.  
Since the new term interferes constructively, the effect is to
increase the decay width.  While the $h\to gg$ mode isn't directly
observable at hadron colliders, this decay width is nevertheless 
important because it
determines the cross-section for Higgs production by gluon fusion;
$\sigma(gg \to h) \propto \Gamma(h \to gg)$, up to soft gluonic
effects.  The $h \to \gamma\gamma$ decay width is similarly modified
by an $\omega$ loop.  In this case, the new contribution interferes
destructively with the SM terms, the dominant parts of which come from
W and Goldstone boson loops.  However, the change in $\Gamma(h \to
\gamma \gamma)$ is very small for any reasonable input parameter
values.  Figure~\ref{brmq} shows the dominant Higgs
decay branching ratios in the model under study.  
Additional NLO corrections to the $h \to gg$ mode were 
included as well, following~\cite{djouadi}.

\begin{figure}[hbt!]
\begin{minipage}[t]{0.5\textwidth}
        \includegraphics[width = \textwidth]{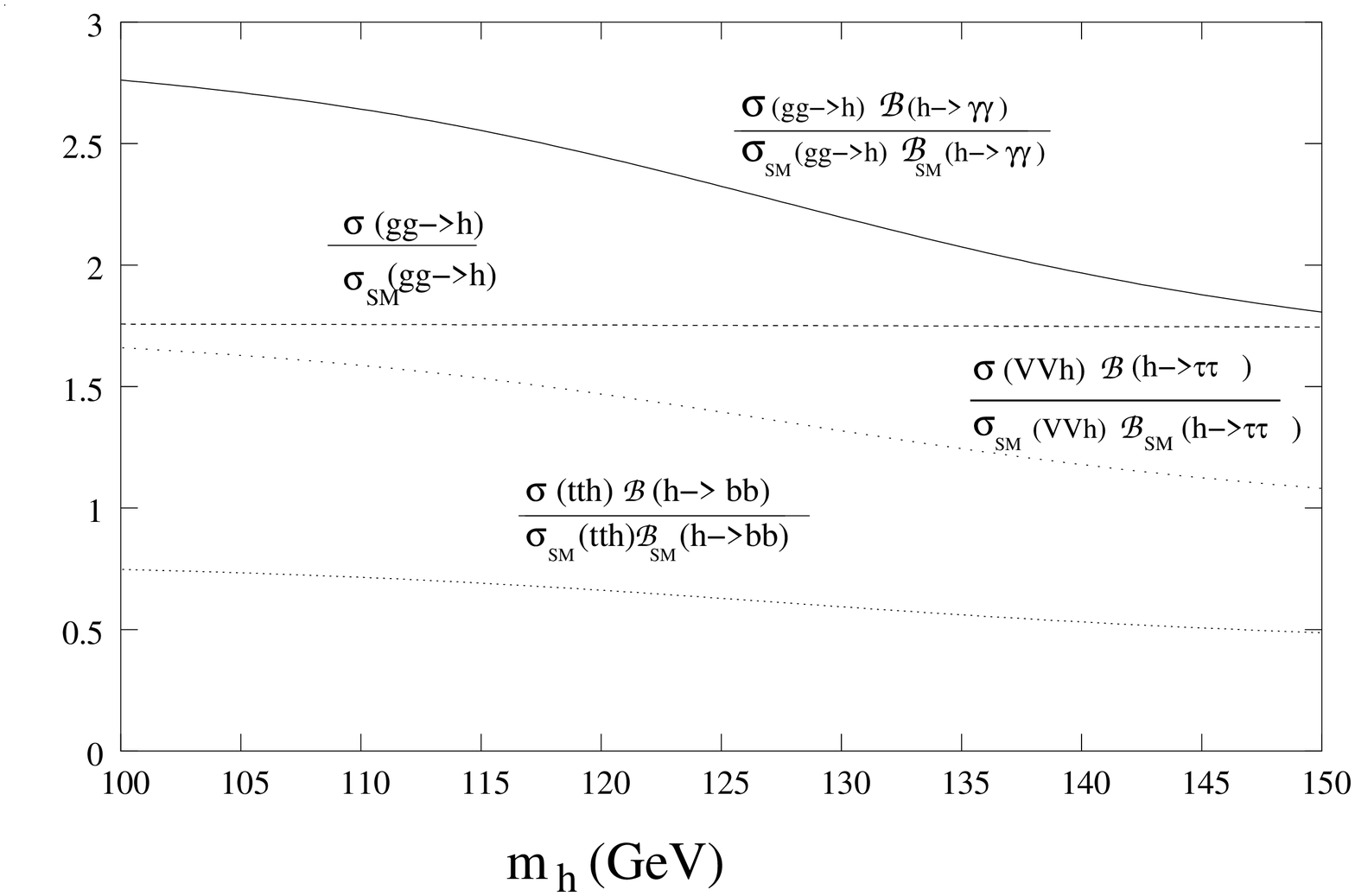}
        \caption{Enhancement of low mass Higgs production and
        detection   
        modes for $m_{\omega} = 250$~GeV, $Y_2/M_1 = 0.7$.
        \label{hdetect}}
\end{minipage}
\phantom{aa}
\begin{minipage}[t]{0.5\textwidth}
        \includegraphics[width = \textwidth]{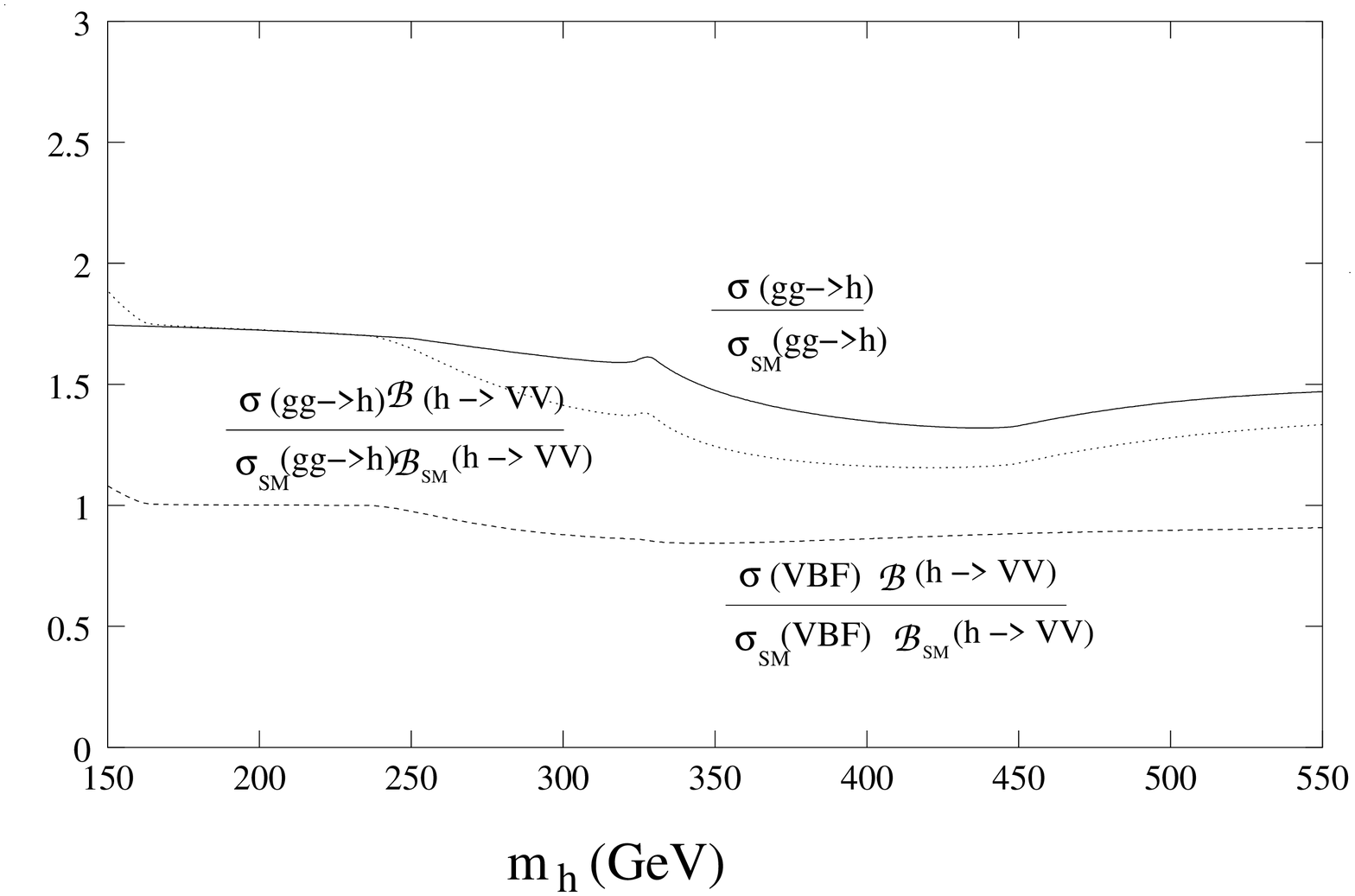}
        \caption{Enhancement of intermediate mass Higgs modes   
        for $m_{\omega} = 250$~GeV, $Y_2/M_1 = 0.7$.}
        \label{hdetect2}
\end{minipage}
\end{figure}

Figures~\ref{hdetect} and~\ref{hdetect2} show the enhancement of a few
Higgs discovery modes relative to the SM, which is mostly due to the
increase in the gluon fusion cross-section.  For low Higgs masses,
$m_h \lesssim 150$~GeV, there is an additional enhancement 
of certain modes as a result of the
decrease in the $h \to \bar{b}b$ branching ratio. 
It should be noted, however, that such low Higgs masses worsen 
the fit to the precision electroweak data in this model.  

If the Higgs mass exceeds $150$~GeV the process
$h \to VV$, where V is a real or virtual vector boson, becomes
the primary Higgs discovery mode at both the Tevatron and the
LHC~\cite{spira1,Carena:2002es}. The inclusive modes are
enhanced by the increase in gluon fusion, while modes in 
which the Higgs is produced by other means,
such as vector boson fusion, are very slightly attenuated due to Higgs
decays into mirror quarks.

   We have checked that including $y_5$ has only a small effect on these results. 
Like the $Zbb$ couplings, the $hbb$ coupling is 
only slightly modified by $y_5$.  The additional mixing due to $y_5$ 
has a larger effect on the mirror quark couplings, increasing the $h\xi\xi$ coupling 
while decreasing that for $h\omega\omega$.  These changes  
precisely cancel each other in Eq.~(\ref{hloop}) to the extent that 
$F_{1/2}(\tau_{\omega}) = F_{1/2}(\tau_{\xi})$, which holds 
to within~$7\%$ for quark masses greater than the Higgs mass.

\subsection{Higgs Searches at the Tevatron and the LHC}
\begin{figure}[hbt!]
\centerline{
\includegraphics[width = 0.7\textwidth]{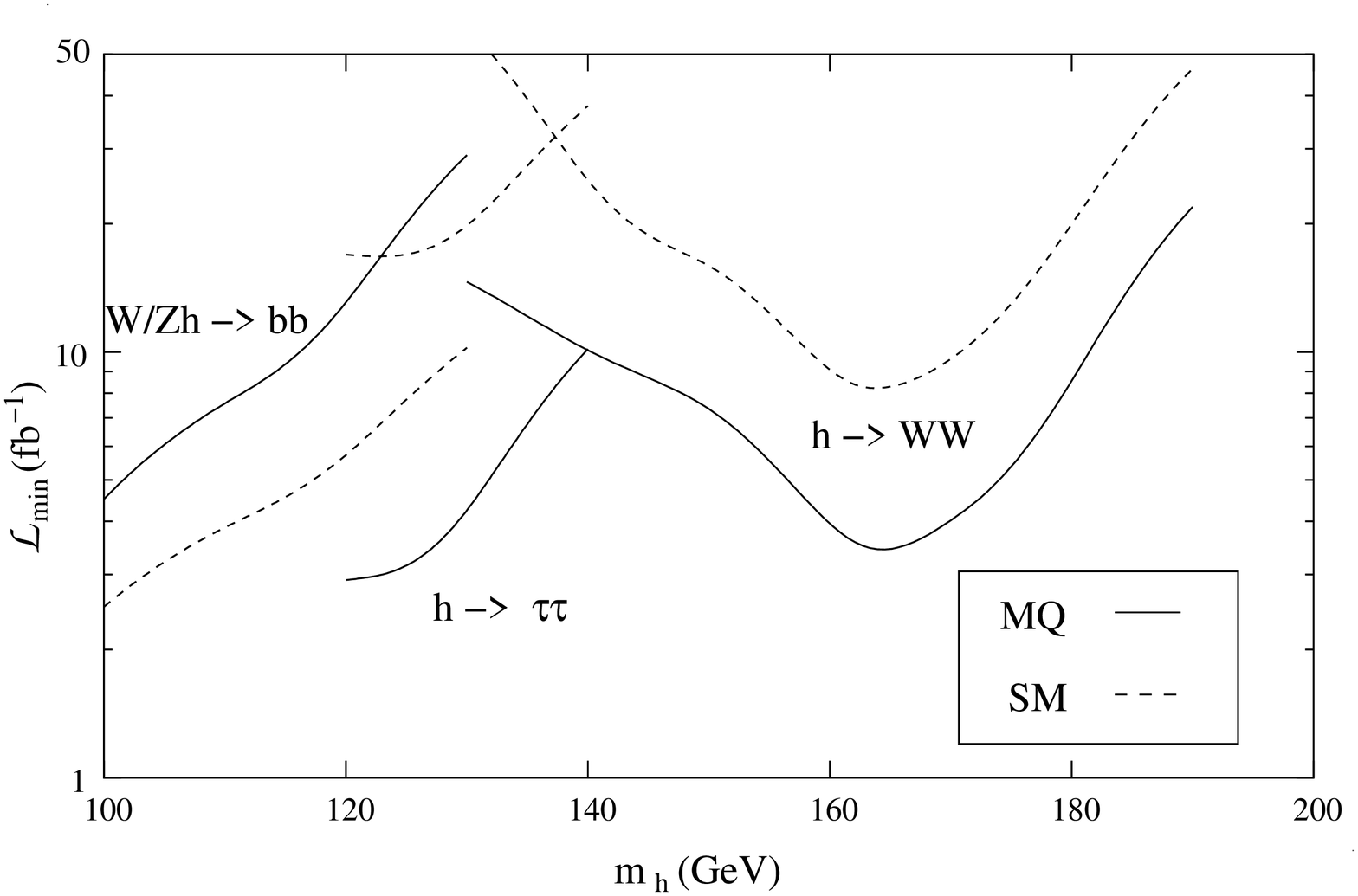}
}
\caption{Minimum luminosity needed for a 3-$\sigma$ Higgs 
signal at the Tevatron for $m_{\omega} = 250$~GeV, $Y_2/M_1 = 0.7$.
\label{tev}}
\end{figure}

\begin{figure}[hbt!]
\centerline{
\includegraphics[width = 0.7\textwidth]{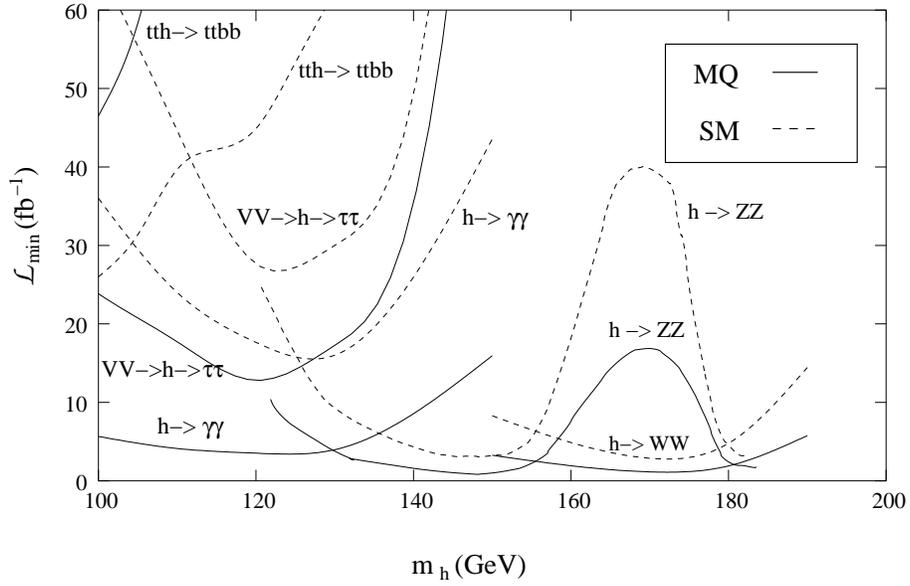}
}
\caption{Minimum luminosity needed for a 5-$\sigma$ Higgs 
discovery at the LHC for $m_{\omega} = 250$~GeV, $Y_2/M_1 = 0.7$.
\label{lhc}}
\end{figure}

   The enhancement of Higgs detection signals decreases the collider
luminosity needed to find the Higgs.  We have translated the above
results into collider units using detector simulation results
from~\cite{Belyaev:2002zz,han,wag} for the Tevatron, 
and~\cite{wag,atlas} for the LHC.
Figure~\ref{tev} shows the minimum luminosity per detector 
(with CDF and D$\emptyset$ data combined) needed for a
3-$\sigma$ signal at the Tevatron.  Figure~\ref{lhc} shows
the luminosity needed for a 5-$\sigma$ discovery at the LHC.
All channels displayed in this plot are for CMS alone
except for the WW and ZZ modes, which are for ATLAS.  
These plots correspond to inclusive searches  
unless stated otherwise:  $VV \to h \to X$ denotes
weak boson fusion channels, while $\bar{t}t h \to X$
and $W/Zh \to X$ denote associated production channels.  
For the inclusive channels,
we have assumed that gluon fusion accounts for $80\%$ of the total
Higgs production.

   The model significantly improves the likelihood of
observing the Higgs at the Tevatron for a Higgs mass between 120
and 180~GeV.  Note that $gg \to h \to \tau^+ \tau^-$ becomes
the dominant discovery channel at the Tevatron collider in the
low Higgs mass region.

The predictions of the model for the LHC are less
dramatic, although the improvement in the $gg \to h \to \gamma\gamma$ 
process will make a light Higgs much easier to see.
For intermediate Higgs masses, the inclusive $h \to WW$ channel becomes
competitive with the $VV \to h \to WW$ channel.
For masses larger than the ones displayed in the Figure~\ref{lhc},
searches for the Higgs at the LHC can proceed via the golden mode
$h \to ZZ$.

\section{Mirror Quark Collider Signals}

If the model is to improve the electroweak fits, the mirror quarks 
must not be too heavy.  In particular, this requires
$m_{\chi} \simlt 250$~GeV, which (using Eq.~(\ref{mass})) implies
$m_{\omega} \simlt 300$~GeV as well.  On the other hand,
$m_{\xi}$ is largely unconstrained.    
These relatively low masses suggest that the Tevatron
may be able to see mirror quarks by the end of Run~II.  
Previous searches for exotic quarks have
concentrated on a possible fourth generation $b'$ quark.  In the
most recent of these, CDF has put a lower bound on the $b'$ mass of
$m_{b'} > 199$~GeV~\cite{affolder}, provided the branching ratio $b'
\to bZ$ is $100\%$.  This bound is relevant to our model as well.

   At the Tevatron, mirror quarks are produced mostly by
$\bar{q}q$ annihilation, with a smaller contribution from gluon
fusion.  Previous calculations of the top-quark pair-production
cross-section apply to mirror quarks as well.  These indicate
$\sigma_{\bar{q}q} \simeq 3.0-0.5$~pb, for $m_{q} = 200 - 300 $~GeV 
at the centre of mass energy $\sqrt{s} = 2.0$~TeV~\cite{berg, kid}.
This is small, but comparable
to the top production cross section in Run~I, $\sigma_{\bar{t}t} = 6.1
\pm 1.1$~pb, where we have averaged the results of D$\emptyset$ 
 and CDF~\cite{6a}.
    
  The up-type $\chi$ quark is most strongly 
constrained in the model.  It decays almost entirely 
by $\chi_R \to b_RW$ due to a large tree-level 
right-handed W coupling, Eq.~(\ref{Jw}). 
This will produce a signature very similar to that of
the top quark.  Indeed, top quark decays present a
nearly irreducible background.  
Searching for the $\chi$ therefore reduces
largely to a counting experiment in which one compares the number of
measured top events to the number expected.  Searches at Run~I of
the Tevatron have already put interesting limits on $m_{\chi}$ since
the top production cross-section measured there agrees
well with SM predictions~\cite{kid} (See, however, the comments 
in~\cite{Sliwa:2002xq}.)

In order to make a quantitative estimate of the present bound on 
a possible sequential top quark,  
we conservatively assume equal acceptance and detection rates for 
both $\chi$ and top events. The fraction of $\chi$ events in the 
top sample will be $40$-$10\%$ for $200<m_{\chi}<250$~GeV.  
(In practice, the detection efficiency for $\chi$'s will 
be slightly better since the b-tagging efficiency improves with jet $P_T$.)
Comparing the Run~I value for the top production cross-section 
with the theoretical prediction~\cite{kid},
we get that the cross-section for any new sequential top quark should
be lower than 2.9 pb, at the 2-$\sigma$ level. This method, based just
on counting, leads to a bound of about 
\begin{equation}
m_{\chi} \simgt 200~{\rm GeV}.
\end{equation}
This in turn implies $m_{\omega} \simgt 250$~GeV, well in excess
of the $b'$ bound from~\cite{affolder}.  

At Run~II, the goal is to determine the top production cross section
to an accuracy of 7--9$\%$~\cite{Savard:2002gc} 
with a few fb$^{-1}$ of data. Assuming that this goal is achieved,
 and considering that the uncertainty is mostly due to
systematic effects (in particular, the proper determination
of the b-tagging efficiency)
and therefore weakly dependent on small luminosity variations, 
the Tevatron Run~II will be sensitive enough to 
rule out the presence of a sequential top quark with a mass smaller than
230~GeV. This will imply an indirect bound on $m_{\omega} > 285$~GeV.
On the other hand, the Tevatron might see evidence of a $\chi$
quark with mass smaller than 220~GeV at the 3-$\sigma$ level.

  That the $\chi \to bW$ vertex is $(V+A)$, Eq.~(\ref{Jw}), makes the $\chi$
somewhat easier to find. This is because the $W^+$'s emitted in $\chi_R
\to b_RW^+$ have positive or zero helicity, whereas those from $t_L
\to b_LW^+$ have negative or zero 
helicity~\footnote{Correspondingly, 
the $W^-$'s emitted in the charge conjugate decays
have negative or zero helicity for a $(V+A)$ vertex, and positive or zero
helicity for a $(V-A)$ vertex.  We shall refer to both positive helicity
$W^+$'s and negative helicity $W^-$'s as ``positive helicity'' 
$W$'s, and so on.}.
Leptons emitted by positive helicity $W$'s tend to be harder than 
those from longitudinal or negative helicity $W$'s.   
Thus, a slightly higher lepton $P_T$ cut will increase
the relative acceptance of $\chi$'s,
although the improvement will be small since the majority 
of $W$'s emitted are longitudinal.  CDF has looked for
positive helicity $W$'s in top decays.  They find a positive helicity
fraction of $\mathcal{F}_+ = 0.11 \pm 0.15$~\cite{whel}, consistent
with both the SM, and a $\chi$ of mass above about $ 200$--250~GeV, for
which we predict a value of $\mathcal{F}_+ \lesssim 0.08$--0.02.  

   Run~II at the Tevatron will also cover part of the 
mass range of the $\omega$ quark by direct searches for this particle.
The strong $\omega b$ mixing leads to tree-level $bZ\omega$
and $bh\omega$ vertices, Eqs.~(\ref{Jz}),(\ref{home}), 
with the same $\mathcal{O}(1)$
flavour-mixing factors.  The dominant decay modes are thus
$\omega_R \to b_R Z$, and $\omega \to b h$ provided the Higgs
isn't heavier than the $\omega$.  Other modes are suppressed by
loops, small flavour-mixing factors, and in the case of $\omega \to
\chi W$, phase space.  Indeed, this decay is forbidden for almost
all of the model parameter space consistent with precision
electroweak data~\cite{0a}.  If the Higgs is heavier than the
$\omega$, the CDF bound applies directly to
$\bar{\omega}\omega \to \bar{b}bZZ$ modes and constrains the
$\omega$ mass to be greater than $199$~GeV.

\begin{figure}[hbt!]
\begin{minipage}[t]{0.5\textwidth}
        \includegraphics[width = \textwidth]{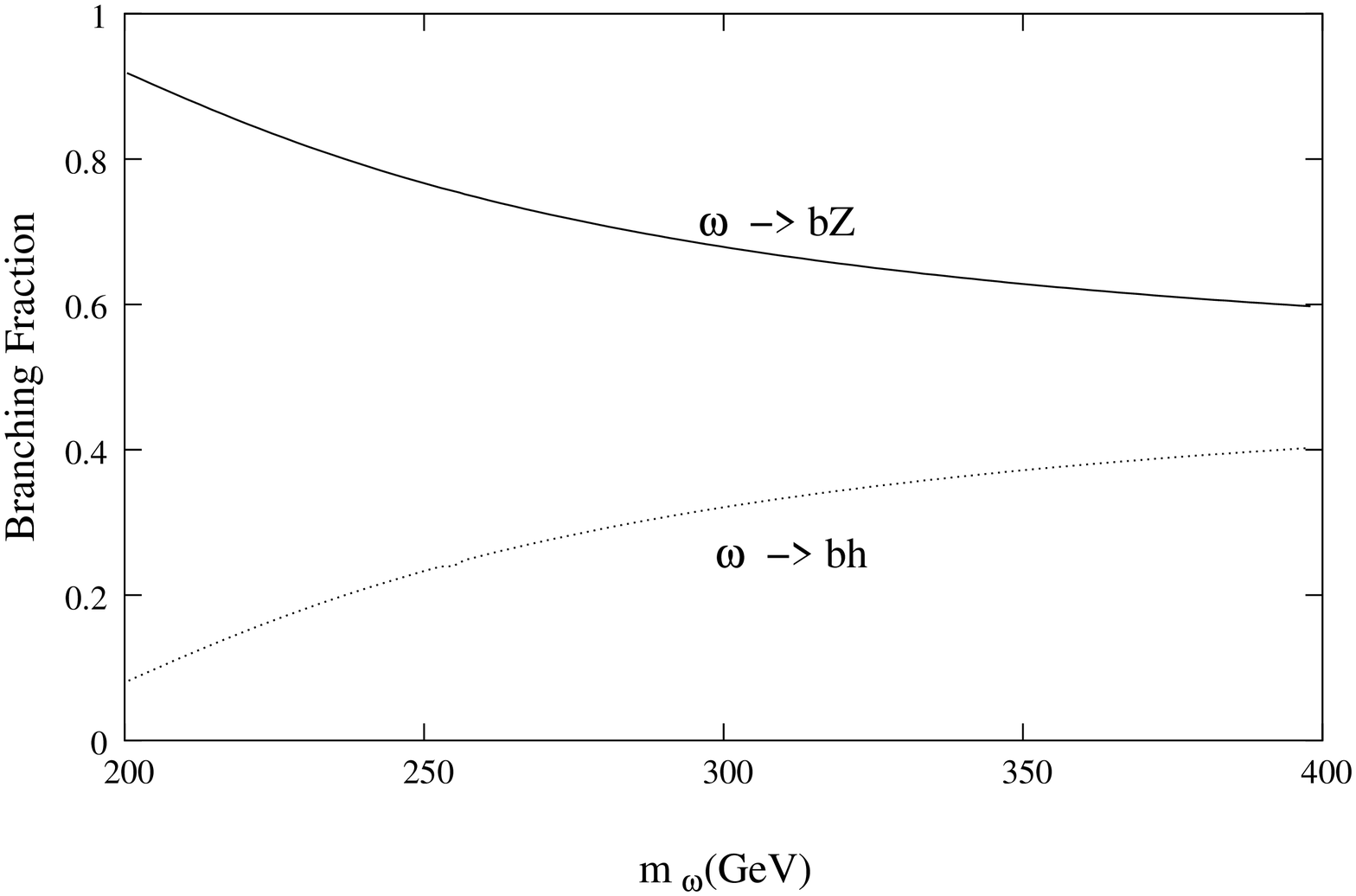}
        \caption{Branching ratios for decays of the $\omega$ quark 
                with $m_h = 170$~GeV.
        \label{ombr}}
\end{minipage}
\phantom{aa}
\begin{minipage}[t]{0.5\textwidth}
        \includegraphics[width = \textwidth]{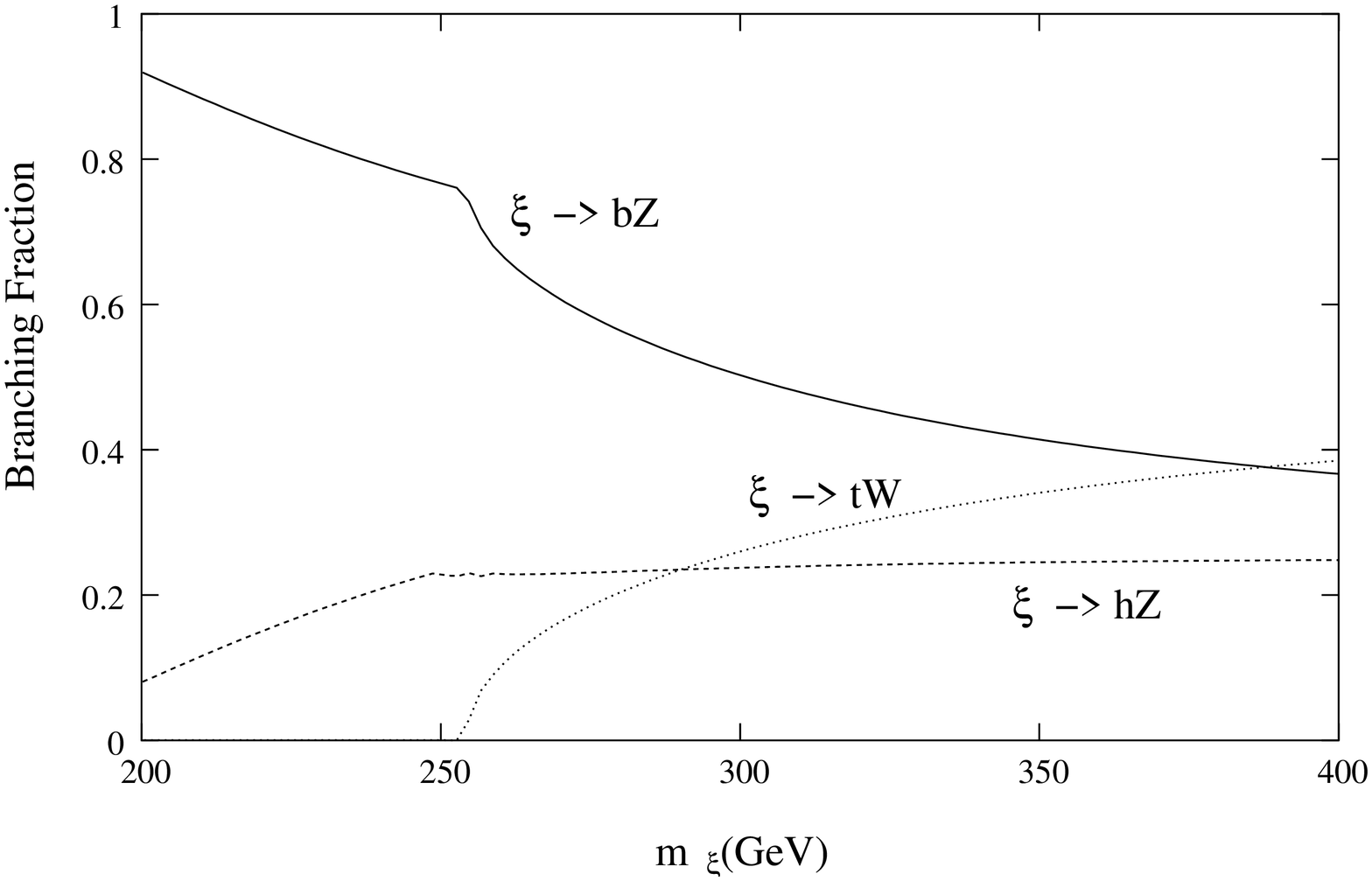}
        \caption{Branching ratios for decays of the $\xi$ quark 
                with $m_h = 170$~GeV (and $y_5=0$).
        \label{xibr}}
\end{minipage}
\end{figure}

Things are more interesting if the Higgs is lighter than the $\omega$.
In this case, the ratio of the decay widths of the Higgs and Z modes
is~\cite{sher} 
\be 
\frac{\Gamma(\omega \to bh)}{\Gamma(\omega \to bZ)}
= \frac{(1 - r_h)^2}{(1-r_Z)^2(1 + 2r_Z)} 
\ee 
where $r_h=(m_h/m_{\omega})^2$, $r_Z = (M_Z/m_{\omega})^2$.
Figure~\ref{ombr} shows the branching ratios for these 
modes for a Higgs mass of $170~$GeV.
In their $b'$ search, CDF looked for $\bar{b}'b' \to \bar{b}bZZ$
events in which one Z decayed into jets while the other decayed into a
pair of high transverse momentum ($P_T$) leptons~\cite{affolder}.  They
only accepted events in which the reconstructed mass of the lepton
pair lay within the range $75$-$105$~GeV and at least two jets were
tagged as b's.  For low Higgs masses, below $150$~GeV, this search
strategy is sensitive to both $\omega\omega \to bZbZ$ and
$\omega\omega \to bZbh$ events, since in the latter, the Higgs decays
predominantly into a $\bar{b}b$ pair which mimics the hadronic decay
of a Z.  This also lends itself to modifying the search strategy to
include four b-tags~\cite{sher}. Even more search strategies become
possible if the Higgs mass exceeds $150$~GeV as favoured by the model.
Such massive Higgs bosons decay mostly into WW pairs (see Figure \ref{brmq})
so for instance, $\omega\omega \to bZbh$ could be distinguished by
looking for events with four jets, at least two of which are b-tagged,
accompanied by a pair of high $P_T$ leptons and large missing $E_T$.

  In order to estimate the possibility of observing an $\omega$ quark at
the Tevatron using the CDF $b'$ search 
strategy, we shall assume that the Higgs 
mass is $m_h = 170$~GeV, and that this search strategy has no sensitivity 
to the $\omega \to bh$ modes and a detection efficiency of 
$13\%$ (as in Run~I for large $m_{b'}$).
The most important background comes from $Z^0$ events 
associated with hadronic jets.  To reduce the background,
one can impose a cut on the total transverse energy  of the
jets~\cite{affolder}.
For $m_{\omega} > 250$ GeV, a cut of $\sum E_T > 150$~GeV will
eliminate most of the background without 
reducing the signal in a significant way.
The number of  observable signal events scales with the 
luminosity and is approximately equal to
\begin{equation}
N_{\omega\bar{\omega}} \simeq 1.5 - 7.0 \;\; {\cal{L}}[{\rm fb}^{-1}] \; , 
\end{equation}
for an $\omega$ mass varying from 300~GeV to 250~GeV.  
Due to the smallness of the background, a simple requirement for 
evidence of a signal is that at least five events be observed.
Therefore the Tevatron 
Run~II should cover the whole mass range of the model, 
$m_{\omega} < 300$ GeV, if the luminosity is above 4 fb~$^{-1}$. 
For a luminosity of 2~fb$^{-1}$,
a signal may be observed up to masses of about 280 GeV.

   The isosinglet $\xi$ decays predominantly by $\xi_L \to b_LZ$,
$\xi_L \to t_LW$, and $\xi \to b h$
(see Eqs.~(\ref{Jz}),(\ref{Jw}) and (\ref{home})).  
All three of these go at
tree level.  Figure \ref{xibr} shows the corresponding branching
ratios for these decays. The $\xi \to bZ$ mode is dominant
for $m_{\xi} \lesssim$ 400~GeV, although the $\xi \to tW$ quickly
becomes important above the $tW$-production threshold~\cite{bigi}.
For relatively light $\xi$'s, of mass less than 250~GeV, the search
strategies are similar to those for the $\omega$.  The CDF $b'$
search is thus sensitive to a light $\xi$.  Their $b'$ mass bound
also applies in this case, although this limit may be weakened somewhat
depending on the mass of the Higgs.

  In the preceding discussion we have taken~$y_5=0$.  We have
checked that including $y_5$ has only a small effect on the $\chi$ 
and $\omega$ decay signatures.  
The most significant change for these decays is 
that the relationship between the masses of the $\chi$ and $\omega$ quarks,
Eq.~(\ref{mass}), need no longer hold, as discussed at the end
of section~\ref{mqmass}.  The branching fractions of the $\omega$
are changed only slightly.  Keeping~$y_5$ has a larger effect 
on~$\xi$ decays, and tends to decrease the branching fraction 
for~$\xi \to tW$ relative to the other modes.

\section{CP-violation in $B^0 \to \phi K_s$ Decays}

The value of the CP-violation parameter $\sin(2\beta)$ measured in
$B^0\to \phi K_s$ decays appears to disagree with the value extracted
from
$B^0 \to J/\psi K_s$ decays:~\cite{hiller, belle, bbar}
\be
        \sin(2\beta) = \left\{ \begin{array}{ll}
                                +0.735 \pm 0.054;&B^0\to J/\psi K_s\\
                                -0.39 \pm 0.41;&B^0\to \phi K_s
                                \end{array} \right.
        \label{s2bdat}
\ee

This discrepancy is particularly interesting because the $B \to \phi
K_s$ mode is loop mediated, making it much more sensitive to new
physics than the $B\to J/\psi K_s$ mode, which goes at tree-level.  We
investigate whether this discrepancy can be explained by the FCNC's
which arise in the model.

It is not $\sin(2\beta)$ that is measured directly, but rather the 
time-dependent CP asymmetry $a_{CP}(t)$.  For decays of the 
$B^0 (= \bar{b}d)$ meson into a CP eigenstate $f$, this is defined to 
be~\cite{fleis}
\bea
a^f_{CP}(t) &:=& 
\frac{ \Gamma(B^0(t) \to f) - 
\Gamma(\bar{B}^0(t) \to f) }{ \Gamma(B^0(t) \to f) + \Gamma(\bar{B}^0(t) \to f) }\\
&=&  \mathcal{C}_f\cos(\Delta Mt) - \mathcal{S}_f\sin(\Delta Mt)\nonumber
\eea
where $\Delta M$ is the mass difference between the mass eigenstates,
 and $\mathcal{S}_f$, $\mathcal{C}_f$ are given by
\be
\begin{array}{cccccc}\mathcal{C}_f &=& \frac{1 - |\xi_f|^2}{1 + |\xi_f|^2},&\phantom{aaa}
        \mathcal{S}_f &=& -\frac{2Im\xi_f}{1 + |\xi_f|^2},
\end{array}
\ee
with $\xi_f$ given by (for the $B^0_d$ system)
\be
   \xi_f \equiv e^{-2i\beta}\frac{A(\bar{B}^0 \to \bar{f})}{A(B^0 \to f)}.
\ee
In the SM, the $\bar{B}\to \phi \bar{K}_s$ amplitude has the form
$\bar{A} = \lambda_tA_0$, where $A_0$ is CP-invariant and
$\lambda_t = V_{CKM}^{ts^*}V_{CKM}^{tb}$.  
The phase of $\lambda_t$ 
is very small~\cite{fleis}, so to a good 
approximation $\mathcal{C}_{\phi K} = 0$, and
$\mathcal{S}_{\phi K}=\sin(2\beta)$.  This result applies to the $B
\to J/\psi K_s$ mode as well.  For this reason, the values of
$\mathcal{S}_f$ measured in the $\phi K_s$ and $J/\psi K_s$ modes are
sometimes quoted as $\sin(2\beta)$, as we have done in ($\ref{s2bdat}$).

   Physics beyond the SM can change the values of $\mathcal{S}_{\phi
K_s}$ and $\mathcal{S}_{J/\psi K_s}$ by adding additional CP-violating
terms to the decay amplitudes.  We have investigated whether
the new tree-level FCNC Z-couplings 
\be 
        J^{\mu}_Z \supset \frac{1}{2\cos\theta_w}\left(\bar{s}_L\gamma^{\mu}V^{sb}_db_L -
\bar{s}_R\gamma^{\mu}\tilde{V}^{sb}_db_R\right) + (h.c.)  
\ee 
can explain the apparent difference between $\mathcal{S}_{\phi K}$ 
and $\mathcal{S}_{J/\psi K_s}$.
To simplify the analysis, we have assumed 
that the mixing of the mirror quarks with the SM quarks 
(other than the b) is very small.  In particular, 
we have neglected all right-handed W couplings and all
FCNC couplings other than those connecting the $b$ and $s$ quarks.
$|V^{sb}|$ and $|\tilde{V}^{sb}|$ will also be treated as small
parameters whose size we will bound below.  These assumptions imply
that the three generation CKM description of flavour mixing in the SM
is correct up to small modifications, and that we can ignore the loop
contributions of mirror quarks to the effective $sb$ vertex.

\subsection{Constraints from Semi-Leptonic Decays}

   The strongest constraints on these couplings come from 
semileptonic $b \to s$ modes~\cite{Barenboim:2000zz}.  At energies 
much below $M_Z$, the SM decay amplitudes can be written as 
the matrix elements of an effective Hamiltonian;
\be
        \mathcal{H}_{eff} = \mathcal{H}_{eff}(b\to s\gamma) 
- \frac{G_F}{\sqrt{2}}\lambda_t(C^{SM}_{9_V}Q_{9_V} + C^{SM}_{10_A}Q_{10_A}),
\ee
where $Q_{9_V}=(\bar{s}b)_{(V-A)}(\bar{l}l)_V,$ and
$Q_{10_A}=(\bar{s}b)_{(V-A)}(\bar{l}l)_A$
are four-quark operators, $\lambda_t = V_{CKM}^{ts^*}V_{CKM}^{tb}$, 
and $l = \mu, e$.  
(See~\cite{buras4} for a definition of $\mathcal{H}_{eff}(b \to s\gamma)$.)  
The FCNC couplings contribute to $Q_{9_V}$ and $Q_{10_A}$, 
and generate the new $(V+A)$ operators
$Q_{9_V}'=(\bar{s}b)_{(V+A)}(\bar{l}l)_V$,
and $Q_{10_A}'=(\bar{s}b)_{(V+A)}(\bar{l}l)_A.$

We neglect the contribution of mirror quarks to
$\mathcal{H}_{eff}(b\to s\gamma)$, since these only arise from loops,
that become negligible for $y_4 = 0$,
contrary to the dominant tree-level effects included in our analysis.
The effective Hamiltonian thus becomes
\be
        \mathcal{H}_{eff} = \mathcal{H}_{eff}(b\to s\gamma) 
- \frac{G_F}{\sqrt{2}}\lambda_t\left(C_9Q_{9_V} + C_{10}Q_{10_A} 
+ C_9'Q_{9_V}' + C_{10}'Q_{10_A}'\right).
\ee

In terms of $\eta := \frac{V_d^{sb}}{\lambda_t}$ and 
$\eta'{:=}-\frac{\tilde{V}_d^{sb}}{\lambda_t}$, 
the Wilson coefficients are now given by
\be
        \begin{array}{cclccl}
        C_{9_V} &=& C_{9_V}^{SM} + (\frac{1}{2} - 
   2\sin^2\theta_w)\eta, & C_{9_V}'&=&(\frac{1}{2} - 2\sin^2\theta_w)\eta',\\
        C_{10_A} &=& C_{10_A}^{SM} - \frac{1}{2}\eta, & C_{10_A}'&=&-\frac{1}{2}\eta'.
        \end{array}
\ee

Since $\frac{1}{2} - 2\sin^2\theta_w \simeq 0.05$, to a good
approximation we need only consider the shift in the $C_{10_A}$
coefficients.  We have examined the effect of modifying the $C_{10_A}$
coefficients on the branching ratios of the inclusive $B \to
X_sl^+l^-$ mode, as well as the two exclusive $B \to Kl^+l^-$ and $B
\to K^*l^+l^-$ decays.

   The constraint on $b\to s$ FCNC's from the $B \to X_sl^+l^-$ mode
has been considered previously (\textit{e.g.}~\cite{buras4, yanir, buch1,handoko}).
We repeat the analysis for this particular model using updated experimental results.
From~\cite{buras4}, modified to include the new
$(V+A)$ operators and neglecting lepton masses, the shift in the branching 
ratio relative to the SM is
\bea
        \Delta\mathcal{B}^{X_s} &=& \mathcal{B} -  \mathcal{B}^{SM}\\
        &=&  \frac{\alpha^2}{8\pi^2f(z)\kappa(z)}\left|\frac{V_{ts}}{V_{cb}}\right|^2
\left(|\tilde{C}_{10_A}|^2 
        + |\tilde{C}_{10_A}'|^2 - |\tilde{C}_{10_A}^{SM}|^2\right)\mathcal{B}
(b \to ce\bar{\nu}),\nonumber
\eea
where $\tilde{C}_i := \frac{2\pi}{\alpha}C_i$, $f(z) = 0.54 \pm 0.04$
is a phase space factor, $\kappa = 0.879 \pm 0.002$ is a QCD
correction, and $\mathcal{B}(b \to ce\bar{\nu}) = 0.109 \pm 0.005$. 
We have also taken $C_{10_A}^{SM} = -\frac{Y_0(x_t)}{\sin^2\theta_w}
\simeq -4.2$, $\alpha^{-1}(m_b) = 129$, and
$\left|\frac{V_{ts}}{V_{cb}}\right|^2 = 1$ in our analysis.  
The measured branching ratio and the SM prediction for this mode are listed in 
Table~\ref{numbers}, as is the 2-$\sigma$ allowed shift in the branching ratio 
based on these values.

\begin{table}[hbt]
\begin{tabular}{|c|c|c|c|}
\hline
Mode&$\mathcal{B}^{exp}( 10^{-6})$&
$\mathcal{B}^{SM}(10^{-6})$&2-$\sigma$ Allowed Range $(10^{-6})$\\
\hline
$B\to X_sl^+l^-$&$6.1^{+1.6}_{-1.4}$~\cite{belle4,
Aubert:2003rv}&
$4.2 \pm 0.7$~\cite{ali5}&-$1.1 < \Delta\mathcal{B}^{X_s} < 5.5$\\
\hline
$B\to Kl^+l^-$&$0.76 ^{+0.19}_{-0.18}$~\cite{belle2, babar1}&
$0.35\pm0.12$~\cite{ali5}&$-0.09 < \Delta\mathcal{B}^{K} < 0.91$\\
\hline
$B\to K^*l^+l^-$&$1.68^{+0.68}_{-0.58} \pm 0.28$~\cite{babar1}&
$1.39 \pm 0.31$~\cite{ali5}&$-1.3 < \Delta\mathcal{B}^{K^*} < 2.1$\\
\hline
\end{tabular}
\caption{Experimental results and SM predictions.  All errors were combined in
quadrature, and the SM predictions were averaged over $\mu$ and $e$ modes.
\label{numbers}}

\end{table}

   The inclusive modes $B\to Kl^+l^-$ and $B\to K^*l^+l^-$ 
have been considered in~\cite{buch1, handoko}.  
The result for the latter mode, neglecting lepton masses, 
is~\cite{buch1}
\bea
        \Delta \mathcal{B}^{K^*} &=&  (4.1^{+1.0}_{-0.7})\times10^{-8}\left(|\tilde{C}_{10_A} 
        - \tilde{C}_{10_A}'|^2 - |\tilde{C}_{10_A}^{SM}|^2\right)\\
        & & \phantom{a} + (0.9^{+0.4}_{-0.2})\times10^{-8}\left(|\tilde{C}_{10_A} + \tilde{C}_{10_A}'|^2 - |\tilde{C}_{10_A}^{SM}|^2\right).\nonumber
\eea
Table~\ref{numbers} lists the experimental results, the theoretical SM prediction,
and the corresponding 2-$\sigma$ allowed range for $\Delta\mathcal{B}^{K*}$.

For the $B \to Kl^+l^-$ mode, the shift in the branching ratio is~\cite{buch1}
\bea
        \Delta \mathcal{B}^{K} = \frac{G_F^2\alpha^2m_B^5}{1536\pi^5}
        \tau_B|\lambda_t|^2I\left(|\tilde{C}_{10_A} 
        + \tilde{C}_{10_A}'|^2 - |\tilde{C}_{10_A}^{SM}|^2\right),
\eea
where $\tau_B = 1.60 \pm 0.05$~ps is 
the total B lifetime, and $I$ is
an integral of form factors.  Explicitly, 
$I = \int^{\hat{s}_1}_{\hat{s}_0}d\hat{s}\lambda_K^{3/2}(\hat{s})f_+^2(\hat{s})$,
where $0 \simeq \frac{4m_l^2}{m_B^2} \le \hat{s} \le \frac{(m_B -
m_K)^2}{m_B^2}$, 
and $\lambda_K(\hat{s}) = 1 + r_K^2 + \hat{s}^2 -
2\hat{s} - 2r_K\hat{s}$ with $r_K = \left(m_K/m_B\right)^2$.  
We have evaluated the integral $I$ numerically 
using the form factors in~\cite{handoko},
and find $I = (0.056^{+0.015}_{-0.09})$.  
Again, table~\ref{numbers} lists the relevant input data.

\begin{figure}[hbt!]
\centerline{
\includegraphics[ width = 0.7\textwidth]{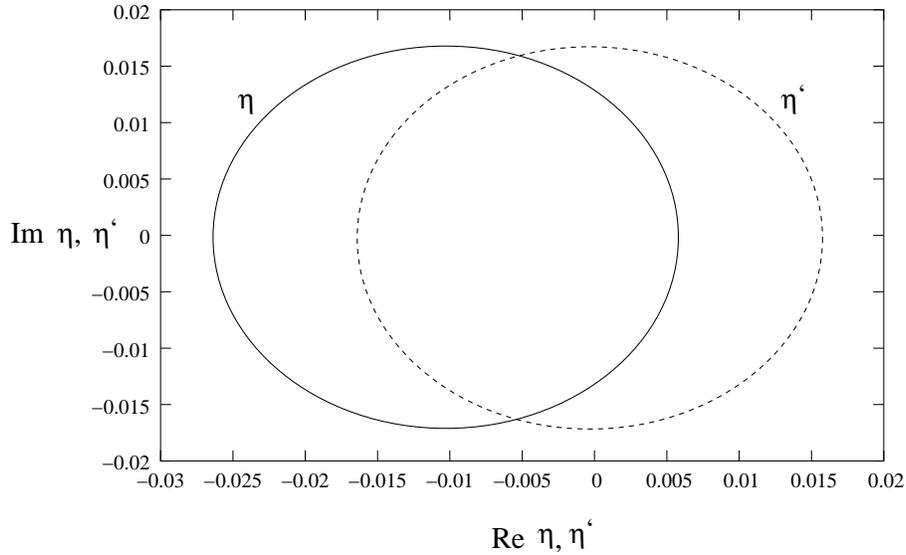}
}
\caption{2-$\sigma$ allowed ranges of $\eta$ and $\eta'$.
\label{scatter}}
\end{figure}

  Figure~\ref{scatter} 
shows the $\eta$ and $\eta'$ values consistent with 
all three semileptonic decay mode constraints taken
at the 2-$\sigma$ level.  Note that points within the two regions
are correlated.  

\subsection{Range of $\mathcal{S}_{\phi K}$}

As for the semi-leptonic modes, the non-leptonic $B \to \phi K_s$
decay amplitude can be written in terms of an effective Hamiltonian.
In the SM, this is given by~\cite{buras2} 
\be \mathcal{H}_{eff}^{SM} =
\frac{G_F}{\sqrt{2}}\left[\lambda_u(C_1Q^{u}_1 + C_2Q^{u}_2) +
\lambda_c(C_1Q^{c}_1 + C_2Q^{c}_2) -
\lambda_t\sum_{i=3}^{10}C_iQ_i\right], \ee 
where $\lambda_q = V^{qs^*}_{CKM}V^{qb}_{CKM}$ are 
products of CKM factors, $Q_i(\mu)$ are
four-fermion operators, $C_i(\mu)$ are the Wilson coefficients, and $\mu$ is the
renormalization scale.\footnote{In writing $\heff$ in this form, we
have made use of $\lambda_u + \lambda_c + \lambda_t \simeq 1$.  This 
is also approximately true when mirror quarks are included.}  The operators 
$Q_1, Q_2$, $Q_3, \ldots Q_6$, and
$Q_7 \ldots Q_{10}$ are the usual SM current-current, QCD penguin, and
electroweak (EW) penguin operators respectively, as defined in~\cite{buras2}.  Note
that the four-quark operators $Q_9$ and $Q_{10}$ are different from the semi-leptonic
operators $Q_{9_V}$ and $Q_{10_A}$ considered above.

   If we include the FCNC couplings from the mirror quarks, 
the tree-level contribution to $\mathcal{H}_{eff}$ at scale $M_W$ is 
\be
        \Delta \mathcal{H}_{eff} = +\frac{G_F}{\sqrt{2}}\lambda_t
\left[\eta(\bar{s}b)_{(V-A)}\sum_q g_{R,L}^q(\bar{q}q)_{(V\pm A)} 
+ \eta'(\bar{s}b)_{(V+A)}\sum_q g_{R,L}^q(\bar{q}q)_{(V\pm A)}\right]
\ee
where the sum runs over $q = u,d,s,c,b$, $g_{R,L}^q = (T_3 -
e_q\sin^2\theta_w)$ is the $Z(\bar{q}q)_{L,R}$ coupling, 
and $\eta = \frac{V_d^{sb}}{\lambda_t}$ and $\eta' = -\frac{\tilde{V}_d^{sb}}{\lambda_t}$ 
are the same as above.  The first operator, 
multiplied by $\eta$, can be written as a linear combination
of $Q_3, \ldots, Q_{10}$.  The second operator, multiplied by $\eta'$,
has no SM counterpart.  We introduce a new ``$(V+A)$'' operator basis
$Q_1', \ldots, Q_{10}'$ related to $Q_1, \ldots, Q_{10}$ by the
interchange $(V-A) \leftrightarrow (V+A)$ wherever these appear.

  We incorporate the new $(V-A)$ operator contribution by modifying the Wilson coefficients at scale $M_W$.  The changes are
\bea
        C_3^{SM}(M_W) &\to& C_3^{SM}(M_W) + \frac{1}{6}\eta,\\
        C_7^{SM}(M_W) &\to& C_7^{SM}(M_W) + \frac{2}{3}\sin^2\theta_w\eta,\nonumber\\
        C_9^{SM}(M_W) &\to& C_9^{SM}(M_W) - \frac{2}{3}(1 - \sin^2\theta_w)\eta.\nonumber
\eea 
For the $(V+A)$ operators, $\left(C_i^{SM}\right)'(M_W) = 0$, 
while the FCNC contribution gives
\bea
        C_5'(M_W) &=& \frac{1}{6}\eta',\\
        C_7'(M_W) &=& -\frac{2}{3}(1 - \sin^2\theta_w)\eta',\nonumber\\
        C_9'(M_W) &=& \frac{2}{3}\sin^2\theta_w\eta',\nonumber
\eea
with all others zero.

   The RG evolution of these operators proceeds much like in the SM
since the $(V-A)$ and $(V+A)$ operators evolve independently.
The anomalous dimension matrix that determines running of both the $(V-A)$ 
and the $(V+A)$ operators is the same as in the SM.  This follows
from our definition of the $(V+A)$ operators, and the fact
that these are renormalized by parity-invariant
gauge interactions.  We calculated the Wilson coefficients at
scale $\mu = 2.5, 5.0$~GeV at one-loop order in both the QCD and QED
corrections using the results of~\cite{buras3}.
  To this order, the corresponding
initial values of the Wilson coefficients are taken at tree-level
in the QCD corrections, although we have included the one-loop electroweak
corrections which give a large contribution to $C_9(M_Z)$.  (This
agrees with the conventions of~\cite{buras2, buras3, benek}.)

   Hadronic matrix elements for the $B \to \phi K_s$ transition at scale 
$\mu = 2.5, 5.0$~GeV 
were estimated using factorization.  Following~\cite{ali1, cheng}, the
amplitude is given by 
\be 
A(\bar{B}\to \phi \bar{K}) =
A_0\lambda_t\left[a_3 + a_4 + a_5 - \frac{1}{2}(a_7 + a_9 +
a_{10}) - \frac{1}{2}(a_7' + a_9' + a_{10}')\right]
\label{bpk}
\ee
where $A_0 = -\sqrt{2}G_Ff_{\phi}m_{\phi}F_1^{BK}(m_{\phi}^2)(\epsilon^*\cdot p_K)$ 
is a CP-invariant product of form factors and constants, 
and the $a_i$ are functions of the Wilson coefficients.  
To leading order, they are~\cite{ali1}: 
$a_{2i-1} = C_{2i-1} + \frac{1}{N_{eff}}C_{2i}$, and 
$a_{2i} = C_{2i} + \frac{1}{N_{eff}}C_{2i-1},$
where $N_{eff}$ is an effective number of colours. 
In writing (\ref{bpk}), we have neglected annihilation contributions. 
While these may be significant, they do not introduce any new weak
phases and are less than~$25\%$ of the SM penguin contribution to 
the amplitude~\cite{cheng}.
We find that the variation of the penguin coefficients 
with $N_{eff}$ is considerably larger than this.

\begin{figure}[!hbt]
\begin{center}
\includegraphics[width=0.8\textwidth]{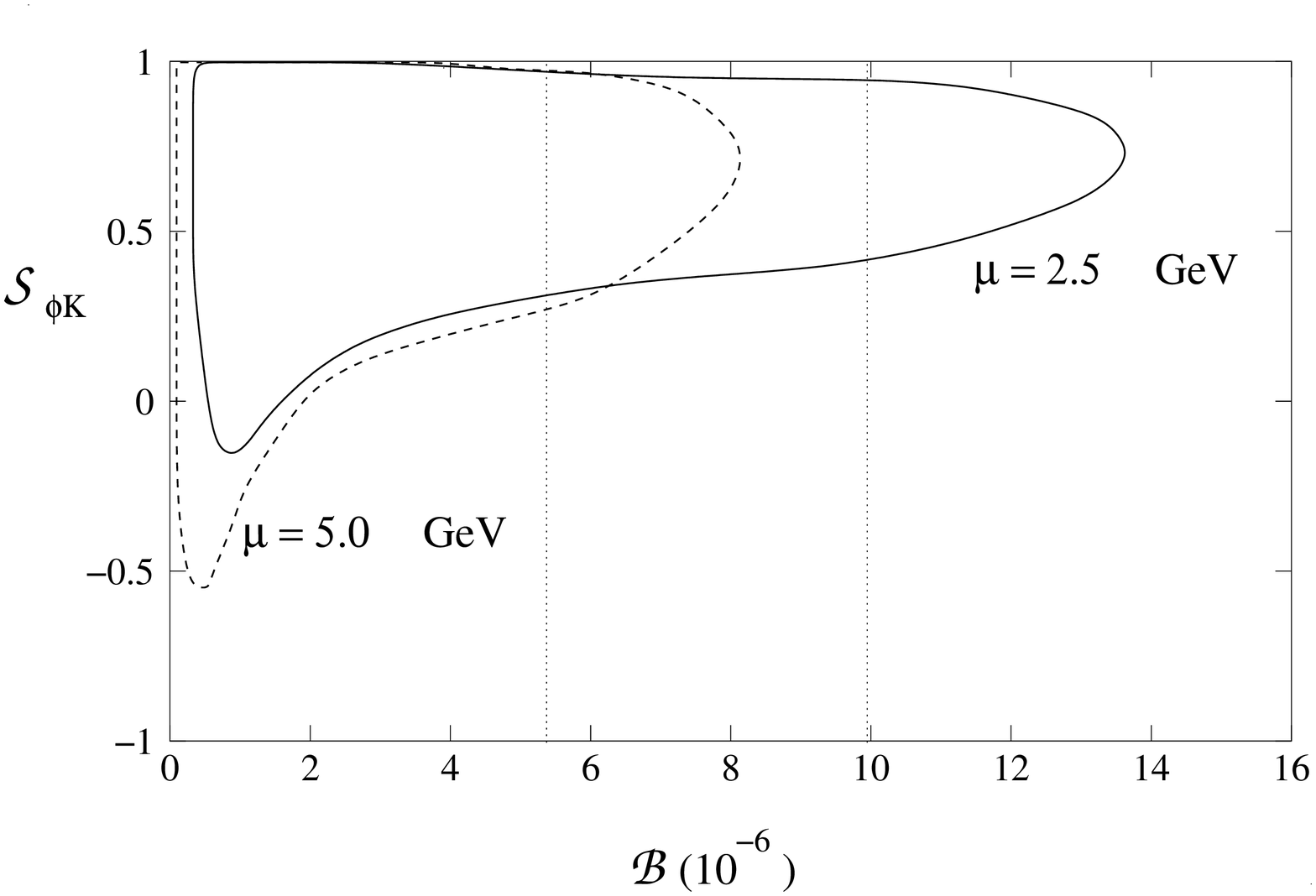}
\end{center}
\caption{Range of $\mathcal{S}_{\phi K}$ accessible by the model 
        plotted against the branching ratio for two choices of
        the renormalization scale.  
        The vertical dotted lines indicate the 2-$\sigma$ allowed 
        region for the branching ratio.
\label{sin2beta}}
\end{figure}

  The $a_i$ coefficients were calculated numerically by taking 
as input the 2-$\sigma$ allowed values of $\eta$ and $\eta'$ from 
the previous section, and running the
Wilson coefficients down to $\mu = 2.5, 5.0$ GeV. 
From these we calculate $\mathcal{S}_{\phi K}$,  
and the $B \to \phi K_s$
branching ratio. This helps to reduce the theoretical uncertainty
due to the sensitivity of the amplitude to variations of $N_{eff}$.
Using (\ref{bpk}) and the input parameters $f_{\phi} = 0.233$ GeV,
$F_1^{BK} = 0.39 \pm 0.03$, $m_{\phi} = 1019.4$~MeV, $m_K = 497.7$~MeV,
$m_B = 5279.3$~MeV, $\tau_{B^0} = 1.54 \pm 0.02$~ps, 
$|\lambda_t| = 0.040 \pm 0.003$, the branching ratio is 
\be
\mathcal{B}(B^0 \to \phi K^0_s) = (5.13 \pm 0.15)\times10^{-3}\left|a_3 + a_4
+ a_5 - \frac{1}{2}(a_7 + a_9 + a_{10}) - \frac{1}{2}(a_7' + a_9' +
a_{10}')\right|^2.  
\ee 
The range of $\mathcal{S}_{\phi K}$ obtained
for $N_{eff} = 2,3,\ldots 10$ is shown in Figure~\ref{sin2beta}, and
is plotted against the branching ratio.  CLEO, BABAR, and BELLE have recently
measured this branching ratio~\cite{Briere:2001ue, Aubert:2003tk, 
unknown:2003jf}, and the average of their results
is $\mathcal{B}(B^0\to \phi K^0) = (7.7 \pm 1.1)\times10^{-6}$.

  We find that a range $0.2 \lesssim \mathcal{S}_{\phi K} \lesssim 1.0$ can be
explained by FCNC's in this model while simultaneously accommodating
semi-leptonic B decay and $B\to \phi K$ branching fraction data at
the 2-$\sigma$ level.  While the shift in 
$\mathcal{S}_{\phi K}$ from FCNC's is not large enough to completely explain 
the current experimental value, it is still significant, and reduces the
discrepancy to below 2-$\sigma$. 
A strong phase $\delta$ in the new physics relative to the SM would only 
decrease the range of $\mathcal{S}_{\phi K}$~\cite{Chiang:2003jn}.
Setting $\delta = \pi$ gives a result similar to that displayed
in Fig.~\ref{sin2beta}. 
The result shows a strong dependence on the renormalization scale, $\mu$,
as well as the effective number colours, $N_{eff}$ due to sensitive 
cancellations between terms in the amplitude.  While this situation would
be improved by adding higher order corrections, we do not
expect such terms to change these general conclusions.

  The above result is more constraining than the one obtained in 
Ref.~\cite{Giri:2003jj}, in which the effect of a
vector-like pair of singlet down quarks on 
$\mathcal{S}_{\phi K}$ was considered.
As we have done here, these authors investigate the range 
of $\mathcal{S}_{\phi K}$
that can be obtained from the $sZb$ vertex that is induced by 
vector-like down quarks.
Our results should reduce to theirs in the limit $\eta' \to 0$, 
which corresponds to considering only the 
flavour mixing effects due to the singlets.  By the same reasoning, 
the range of $\mathcal{S}_{\phi K}$ should be greater in the present model
since even more flavour mixing is possible.
Instead, these authors find a larger range for $\mathcal{S}_{\phi K}$ 
than we have obtained. One of the possible sources of discrepancy
between our analysis and the
one in Ref.~\cite{Giri:2003jj} resides in our use of a more
stringent quantitative analysis of
the constraints coming from the semileptonic B-decays.  
Another source appears to be 
their inclusion of the ``colour-suppressed'' operators 
$(\bar{s}_{\beta}b_{\alpha})_{(V-A)}(\bar{s}_{\alpha}s_{\beta})_{(V\pm A)}$
at tree-level.  While such operators do arise
from QCD corrections to the $sZb$ vertex,  
all tree-level effects may be described by the ``colour allowed'' operators  
$(\bar{s}_{\beta}b_{\beta})_{(V-A)}(\bar{s}_{\alpha}s_{\alpha})_{(V\pm A)}$.

\section{Conclusion}

  We have investigated the phenomenological properties of 
  Beautiful mirrors, an extension
  of the Standard Model consisting of additional vector-like
  ``mirror'' quarks with the same quantum numbers as  
  the $SU(2)$ quark doublet and
  down quark singlet in the Standard Model. These exotic
  quarks mix with the bottom quark resulting in a modified value of
  the right-handed bottom quark coupling to the $Z$ gauge boson, in
  agreement with indications coming from the precision electroweak data.
  A good fit to the precision electroweak data also demands that the additional
  quarks have masses lower than about $300$~GeV implying a rich phenomenology
  at the Tevatron and LHC Colliders, as well as a possible impact on the 
  CP-violating observables measured at the B-factories. In addition, 
  the unification of gauge couplings is greatly improved within the model.   

  In this article we have provided a detailed analysis of the question of
  gauge coupling unification. 
    We find that the gauge couplings unify at $M_G = (2.80
  \pm 0.15)\times 10^{16}$~GeV.  Perturbative consistency and stability
  of the model restrict the possible values of the masses of the Higgs
  and the mirror quarks.  The allowed range, $m_h = 170 \pm 10$~GeV, 
  overlaps with the range of
  values of these parameters which give the best fit to precision
  electroweak data~\cite{0a}. 

  Flavour mixing due to the mirror quarks leads
  to right-handed Z couplings, a very small
  loss of unitarity of the CKM matrix, and FCNCs.  
  The flavour mixing also modifies the coupling 
  of the $b$ and $\omega$ quarks to
  the Higgs, while the couplings of the other quarks to the Higgs
  are not changed significantly.  This has some interesting
  implications for Higgs searches at the Tevatron. In particular,
  the required luminosity for a Tevatron or LHC Higgs discovery in the 
  $WW$ decay mode, as well as in the $\tau^+\tau^-$ mode at 
  the Tevatron and the $\gamma \gamma$ mode at the LHC,
  is greatly reduced within this model. 
  We have analyzed the search for mirror vector quarks
  at the Tevatron collider, and have found that Run~II with
  a total integrated luminosity of about 4~fb$^{-1}$ will be able to
  test all of the mirror quark mass range consistent with electroweak
  precision data.
  Finally, the $b\to s$ FCNCs which arise in this model can help
  explain the discrepancy between the values of $\sin(2\beta)$ 
  measured in the $B\to \phi K$ and $B \to J/\psi K$ decays.

~\\
{\bf Acknowledgements} The authors would like to thank M.~Carena,
Z.~Chacko, C.W.~Chiang, D.~Choudhury, D.E.~Kaplan. T.~Le~Compte,
R.~Sundrum and T.M.P.~Tait for interesting
comments and suggestions. C.W. would like to thank the Aspen Center
for Physics, where part of this work has been done.
Work supported in part by the US DOE, Div.\ of HEP,
Contract W-31-109-ENG-38. 
~\\
~\\
{\bf Note added :}    As this work was being completed,
Ref.~\cite{Atwood:2003tg} appeared,
in which the effect of tree-level $(V \pm A)$ sZb couplings on
$\mathcal{S}_{\phi K}$ was calculated.  The modifications to the
electroweak scale Wilson coefficients as well as the final result in 
this paper agree with our analysis. \\
After this work appeared, revised experimental values of 
$\mathcal{S}_{\phi K}$ were reported by the Babar and Belle
experiments. The new world average is still more than two standard
deviations away from $\mathcal{S}_{J/\psi K_S}$, although there is also
a 2.1 $\sigma$ discrepancy between the two experimental 
results~\cite{Leppho}.\\
A related work,~\cite{Andre:2003wc} has also recently appeared,
in which the search for $Q=-1/3$ vector-like quarks is investigated.
The conclusions of these authors for the Tevatron are similar to ours.

\end{document}